\documentclass[12pt]{article}
\usepackage{amssymb}
\usepackage{graphics}

\def\section#1{\addtocounter{section}{1}
    \setcounter{subsection}{0}
    \setcounter{subsubsection}{0}
    \setcounter{equation}{0}\vskip8mm
    \begin{center}{\bf\large \thesection~#1}\end{center}}
\def\subsection#1{\addtocounter{subsection}{1}
    \setcounter{subsubsection}{0}
    \vskip6mm\noindent{\sc\large \thesubsection~#1}\vskip4mm}
\def\subsubsection#1{\addtocounter{subsubsection}{1}
    \vskip6mm\noindent{\bf \thesubsubsection~#1}\vskip4mm}
\def\paragraph#1{\vskip6mm\noindent\underline{\bf #1}\vskip3mm}
\def\thebibliography#1{\list
 {[\arabic{enumi}]}{\settowidth\labelwidth{[#1]}\leftmargin\labelwidth
  \advance\leftmargin\labelsep
  \usecounter{enumi}}
  \def\newblock{\hskip .11em plus .33em minus .07em}
  \sloppy\clubpenalty4000\widowpenalty4000
  \sfcode`\.=1000\relax}

\renewcommand{\theequation}{\arabic{section}.\arabic{equation}}
\newcommand{\nn}{\nonumber}
\newcommand{\ts}{\textstyle}
\newcommand{\ds}{\displaystyle}
\newcommand{\Z}{{\mathbb Z}}
\newcommand{\R}{{\mathbb R}}
\newcommand{\Dbar}{\overline{\rm D}}
\newcommand{\mathW}{{\mathbb W}}
\newcommand{\mathP}{{\mathbb P}}
\newcommand{\mathO}{{\mathcal O}}
\newcommand{\mathZ}{{\mathbb Z}}
\newcommand{\mathC}{{\mathbb C}}
\newcommand{\mathR}{{\mathbb R}}
\newcommand{\w}{\wedge}
\newcommand{\p}{\partial}
\newcommand{\n}{{\mathcal N}}
\newcommand{\Jacobi}{{\rm J}}
\newcommand{\beq}{\begin{equation}}
\newcommand{\eeq}{\end{equation}}

\def\+{e^{\frac{\, i \alpha \tau_3}{2}}}
\def\-{e^{\frac{-i \alpha \tau_3}{2}}}

\begin{document}

\baselineskip=6mm
\renewcommand{\thefootnote}{\fnsymbol{footnote}}
\setcounter{footnote}{0}
\setcounter{section}{0}
\pagestyle{empty}

\hfill\vbox{\hbox{hep-th/0501195} }

\vskip30mm

\begin{center}
 {\large\bf $\bf G_2$ Holonomy, Mirror Symmetry and Phases of N=1 SYM}
\end{center}

\vskip10mm

\begin{center}
    Kazuo Hosomichi and David C. Page
 \\ \vskip3mm
{\it Department of Physics, University of Toronto,\\
     60 St.George Street, Toronto, Ontario M5S 1A7, Canada} \vskip3mm
\end{center}

\vskip8mm

\begin{center}{\bf Abstract}\end{center}

\vskip4mm

We study the phase structure of four-dimensional ${\cal N}=1$
super Yang-Mills theories realized on D6-branes wrapping the
$\mathR\mathP^3$ of a $\mathZ_2$ orbifold of the deformed
conifold. The non-trivial fundamental group of $\mathR\mathP^3$
allows for the gauge group to be broken to various product groups
by $\mathZ_2$ Wilson lines. We study the classical moduli space of
theories in various pictures related by dualities including an
M-theory lift. The quantum moduli space is analyzed in a dual IIB
theory, where a complex curve contained in the target space plays
a key role. We find that the quantum moduli space is made up of
several branches, characterized by the presence or absence of a
low energy $U(1)$ gauge symmetry, which are connected at points of
monopole condensation. The resulting picture of the quantum moduli
space shows how the various gauge theories with different product
gauge groups are connected to one another.

\vspace*{\fill}
\noindent January~~2005
\setcounter{page}{0}
\newpage


\setcounter{footnote}{0}
\setcounter{section}{0}
\pagestyle{plain}
\renewcommand{\thesection}{\arabic{section}.}
\renewcommand{\thesubsection}{\arabic{section}.\arabic{subsection}.}
\renewcommand{\theequation}{\arabic{section}.\arabic{equation}}
\renewcommand{\thefootnote}{\arabic{footnote}}
\setcounter{footnote}{0}

\section{Introduction}

Understanding the dynamics of string compactifications to four
dimensions with ${\cal N}=1$ supersymmetry is one of the most
important challenges confronting string theory. A better
understanding of this subject is necessary in order to learn more
about the vacuum structure of string theory and will be crucial
for connecting to real world physics. One problem of fundamental
interest is to understand the structure of the moduli space of
${\cal N} =1$ string backgrounds and we shall study a particular
example of this in the current work.

One fruitful example, which has been studied by a number of
authors, is the case of the geometric transition for type II
strings compactified on the conifold in the presence of branes and
flux. This study was initiated in \cite{Vafa:2000wi} (following
earlier work on topological strings  \cite{Gopakumar:1998ki}) , in
which the open-closed string duality involving D6-branes wrapping
the minimal $S^3$ of the deformed conifold
\begin{equation}
 z_1^2+z_2^2+z_3^2+z_4^2 = \mu,
\label{defc}
\end{equation}
was related to the dynamics of ${\cal N}=1$ pure SYM theory. A
large number of D6-branes bring the geometry into a resolved
conifold with RR2-form flux. String theory relates the SYM scale
$\Lambda$ with the volume of $S^3$ and the gaugino condensate $S$
with the volume of the blown-up $S^2$, and the relation between
the deformed and resolved conifolds gives a precise relation
between $\Lambda$ and $S$.

This smooth geometric transition of the conifold was understood
from the M-theory viewpoint in
\cite{Acharya:2000gb,Atiyah:2000zz,Atiyah:2001qf}. For $SU(N)$ SYM
theory, it was shown that the compactification manifold should be
a $\mathZ_N$ orbifold of $\mathR^4\times S^3$ with metric of $G_2$
holonomy,
\begin{equation}
 ds^2 ~=~
 \frac{dr^2}{1-(r_0/r)^3}+
 \frac{r^2}{36}(1-(r_0/r)^3)
 \sum_{a=1}^3(\sigma_a+\tilde\sigma_a)^2
 +\frac{r^2}{12}\sum_{a=1}^3(\sigma_a-\tilde\sigma_a)^2.
\end{equation}
Here $r_0$ denotes the unique non-normalizable deformation of the
background, and $\sigma_a$ and $\tilde\sigma_a$ are two sets of
$SU(2)$ Maurer-Cartan 1-forms. As was discussed in
\cite{Atiyah:2001qf}, the classical moduli space of $G_2$ holonomy
manifolds $\mathR^4\times S^3$ consists of three half-lines
emenating from a point. This can be understood as follows. We
regard the space as a deformation of a cone over
\begin{equation} \label{equiv}
 S^3\times S^3 ~\cong~ SU(2)^3/SU(2) ~\equiv~
 \{(g_1,g_2,g_3)\sim(g_1h,g_2h,g_3h)~|~g_{1,2,3},h\in SU(2)\} \, .
\end{equation}
The three branches of classical moduli space correspond to the
three choices of $S^3$ -- $g_1, g_2$  or $g_3$ -- to fill in in
order to make the $\mathR^4$. On one branch, the $\mathZ_N$ acts
on the $\mathR^4$ part and has an $S^3$ fixed point locus, leading
to $SU(N)$ gauge dynamics, whereas on the other two branches, the
$\mathZ_N$ acts freely and there is no gauge dynamics.

Including the modulus corresponding to the M-theory three-form
potential through $S^3$, the moduli space is complex
one-dimensional and classically is given by three semi-infinite
cylinders meeting at a point. Quantum effects turn this into a
single smooth quantum moduli space which is a Riemann sphere with
three semi-classical (large volume) points. This quantum moduli
space was shown to encode the dynamics of the SYM theory (with
some higher order corrections) such as the relation between
$\Lambda$ and $S$ mentioned above. Note that the Riemann sphere
contains the vacua corresponding to different $\Lambda$, so it
should better be called the moduli space of vacua and theories.

Through the analysis of various orbifold groups,
\cite{Atiyah:2001qf} also showed that  two gauge theories with
different microscopic gauge groups can sometimes appear on the
same string theory moduli space. One example is the case of the
orbifold of $\mathR^4\times S^3$ by the dihedral group $D_{4+n}$,
where two SYM theories with gauge groups $SO(2n+8)$ and $Sp(n)$
are connected on the same moduli space. As another example,
\cite{Friedmann:2002ct} studied  more complicated orbifold groups
of the form $\Gamma=\Gamma_1\times\Gamma_2\times\Gamma_3$ which,
among other things, give rise to the possibility for the gauge
groups to be broken by discrete Wilson lines (see also
\cite{Ita:2002ws}).

An example of this kind is the orbifold
$(\mathR^4/\mathZ_N)\times(S^3/\mathZ_2)$ corresponding to $N$
D6-branes on a $\mathZ_2$ orbifold of conifold, and this is the
model which we will focus on in this paper. Since one has a choice
of $\mathZ_2$ Wilson line $\pi_1(S^3/\mathZ_2)\to SU(N)$, there
arise backgrounds corresponding to various classical gauge groups
$U(1)\times SU(N_+)\times SU(N_-)$ with $N_++N_-=N$. One can then
ask how all these backgrounds are put together to form a quantum
moduli space. A recent related work \cite{Cachazo:2002zk,
Cachazo:2003yc} has obtained, in a different setting, smooth
moduli spaces that interpolate between vacua of SYM theories with
different classical gauge groups.

It has recently been conjectured that the $U(N)$ Chern-Simons
theory on $\mathR\mathP^3$ is dual to the closed topological
A-model on a non-compact Calabi-Yau manifold ${\cal
O}(K)\to\mathP^1\times\mathP^1$ \cite{Vafa:2000wi,Aganagic:2002wv,
Okuda:2004rg}, and the corresponding large $N$ duality is expected
to hold also in superstring theory. On the Chern-Simons side,
there is a choice of $\mathZ_2$ Wilson line which breaks the gauge
group to product groups. The numbers $N_+,N_-$ appear on the
closed string side as the sizes of two $\mathP^1$'s and are also
proportional to the RR2-form flux piercing through them. By
analogy with the story for the conifold, one expects that this
resolved geometry describes the low energy dynamics of SYM
theories with these product gauge groups.

The aim of this paper is to obtain the phase structure of
four-dimensional string compactifications which are on the same
${\cal N}=1$ moduli space as the background with $N$ D6-branes
wrapping the minimal $\mathR\mathP^3$ of a $\mathZ_2$ orbifold of
the deformed conifold. We will first study the classical moduli
space by listing the classical solutions or brane configurations
with the same asymptotic boundary conditions in a given string or
M-theory setup, and then move on to a study of the quantum moduli
space.

The analysis will be made from several frameworks related by a
chain of dualities. As was shown in \cite{Aganagic:2001jm,
Atiyah:2001qf}, D6-branes wrapped on the $S^3$ of the deformed
conifold are dual, via an M-theory lift, to a configuration of a
Lagrangian D6-branes on an orbifold of $\mathC^3$, which admits a
gauged linear sigma model description. This also has a mirror IIB
description in terms of D5-branes on a non-compact Calabi-Yau. We
will find the IIB framework to be the most powerful for analyzing
the quantum moduli space. Finally, there is a duality to
three-dimensional webs of 5-branes in IIB string theory which
offers a particularly useful description of the classical moduli
space.

~

This paper is organized as follows. In section 2 we describe our
problem and discuss the classical moduli space in the original
type IIA setting. It is up-lifted to M-theory in section 3, where
we explain the various $G_2$ holonomy spaces corresponding to
branches of classical moduli space. In addition to orbifolds of
the $\mathR^4\times S^3$ geometry explained above, we will need
different $G_2$ holonomy solutions with the same asymptotics and a
minimal $T^{p,q}$ at the center. In section 4 we dimensionally
reduce to another IIA picture and obtain a description of the
classical moduli space in terms of a gauged linear sigma model.
Various $G_2$ holonomy manifolds are mapped to configurations of a
D6-brane in different partial blow-ups of a toric orbifold
$\mathC^3/(\mathZ_N\times\mathZ_2)$. Also, in sections 3,4 we
explain another type IIB dual in which these classical vacua are
described as three-dimensional webs of 5-branes with fixed
asymptotics. We will make use of these webs to prove that there
are no further classical vacua with the required asymptotics.
Finally, in section 5 we take the mirror of the GLSM and move to a
IIB theory on a non-compact Calabi-Yau space containing a complex
curve $\Sigma$. The D6-brane of the IIA theory turns into a
non-compact D5-brane intersecting with $\Sigma$ at a point. The
presence of this brane generates a superpotential $W$ for the
closed string moduli. By a careful analysis of $\Sigma$ and $W$ we
will obtain the branch structure of the quantum moduli space. This
moduli space will turn out to interpolate between various SYM
theories with different product gauge groups.

\section{Classical phases of the type IIA geometry}
We are interested in a $\mathZ_2$ orbifold of the deformed
conifold (\ref{defc}), where $\mathZ_2$ acts as $z_i \to -z_i$.
Acting on the deformed conifold, the $\mathZ_2$ is fixed point
free and the minimal $S^3$ becomes an $\mathR\mathP^3$. Indeed the
orbifold space is topologically (and symplectically) $T^*
\mathR\mathP^3$.

We wish to consider $N$ D6-branes wrapping the base $\R \mathP^3$.
Naively we would expect to find a $U(N)$ gauge theory living on
the worldvolume of the D-branes. On lifting to M-theory, we find
an $SU(N)$ gauge group as discussed in \cite{Brandhuber:2001yi}
and reviewed in the next section. We can also see this within the
IIA setup as the resolved conifold does not have a normalizable
harmonic two-form and thus the $U(1)$ is missing after geometric
transition. We shall assume that the $SU(N)$ gauge group is
correct although we do not understand the mechanism for removing
the extra $U(1)$ in the D6-brane theory.

Note that this $SU(N)$ gauge theory has many more vacua than the
case without orbifold, since we can include Wilson lines
$\pi_1(\R\mathP^3) \cong \Z_2 \rightarrow SU(N)$ which break the
gauge group to $SU(N_+) \times SU(N_-) \times U(1)$. If $N_-$
labels the number of $(-1)$ eigenvalues of the Wilson line, then
$N_-$ is necessarily even, since the gauge group is $SU(N)$ not
$U(N)$. All this suggests a richer structure for the moduli space
than in the case of the conifold.

Other semi-classical limits of moduli space should be described by
geometries which are asymptotic to a cone over $T^{1,1}/\Z_2$ with
$N$ units of RR-flux through the $S^2$. One possibility is the
$\Z_2$ orbifold of the resolved conifold. In this case, the $\Z_2$
has a $\mathP^1$ of fixed points and we can consider blowing up
the singularity to form a space with two non-trivial $\mathP^1$'s.
The blown up space is $\mathO(K) \rightarrow \mathP^1 \times
\mathP^1$.

This geometry has two K\"ahler parameters. The first is related to
an overall rescaling of the metric whilst the other controls the
relative size of the two $\mathP^1$'s. The overall rescaling is a
non-normalizable parameter labelling different points on the
moduli space of theories. The relative rescaling of the two
$\mathP^1$'s is normalizable and corresponds to a dynamical field,
which gets frozen by a flux superpotential. The value at which it
gets frozen depends on the background RR-flux.

In fact it is possible to be completely explicit here. The
Calabi-Yau metric on $\mathO(K) \rightarrow \mathP^1 \times
\mathP^1$ was found in \cite{PandoZayas:2001iw}. It is\footnote
{
Although the metric is written in terms of seven independent
coordinates, it depends on $\psi,\tilde\psi$ (introduced below)
only through $\psi-\tilde\psi$.
}: \beq
\label{p1} ds^2 = \kappa^{-1} (\rho) d \rho^2 + \frac{1}{9}
\kappa(\rho) \rho^2 (\sigma_3 - \tilde{\sigma}_3)^2 + \frac{1}{6}
\rho^2 (\sigma_1^2 + \sigma_2^2) + \frac{1}{6}(\rho^2 +
6a^2)(\tilde\sigma_1^2 + \tilde\sigma_2^2) \, ,\eeq where \beq
\kappa(\rho) = \frac{1 + \frac{9 a^2}{\rho^2}-
\frac{b^6}{\rho^6}}{1 + \frac{6a^2}{\rho^2}}  \eeq and $a$ and $b$
are parameters. Varying $a$ is a non-normalizable deformation
whilst varying $b$ is normalizable.

As in the introduction $\sigma_a,\tilde\sigma_a$ are two sets of
$SU(2)$ Maurer-Cartan forms. Explicitly, $X^{-1}dX = \frac
i2\tau_a\sigma_a$ where $\tau_a$ are Pauli's matrices,
\begin{equation}
  X = \left(\begin{array}{rr}
      \cos\frac\theta2e^{ \frac{i}{2}(\psi+\phi)} &
     -\sin\frac\theta2e^{-\frac{i}{2}(\psi-\phi)} \\
      \sin\frac\theta2e^{ \frac{i}{2}(\psi-\phi)} &
      \cos\frac\theta2e^{-\frac{i}{2}(\psi+\phi)}
      \end{array}\right)~~~~
 (0\le \psi < 4\pi,~~~
  0\le \phi < 2\pi,~~~
  0\le \theta < \pi)
\end{equation} and $\tilde{X}(\tilde{\psi},\tilde{\theta},\tilde{\phi})$ and
$\tilde{\sigma}_a$ are defined similarly.
The radial coordinate $\rho \in [\rho_0 , \infty)$ where $\rho_0$
is defined as the positive value of $\rho$ at which $\kappa(\rho)$
vanishes. In order to remove a conical singularity at this locus,
it is necessary to orbifold by $\Z_2$ so that $\check{\psi} = \psi
- \tilde{\psi}$ has length $2 \pi$. At $\rho \rightarrow \infty$
the metric approaches the standard metric on (the $\Z_2$ orbifold
of) the conifold.

The K\"ahler form corresponding to this choice of metric is: \beq
\omega_0 = \frac{1}{3}\rho \, d\rho \w (\sigma_3 -
\tilde{\sigma}_3) + \frac{1}{6}\rho^2 \sigma_1 \w \sigma_2 -
\frac{1}{6} (\rho^2 + 6 a^2) \tilde\sigma_1 \w \tilde\sigma_2 \, .
\eeq We should allow fluctuations of the K\"ahler class by
elements of normalizable second cohomology. There is one such
class, which is generated by the two-form: \beq \label{omega1}
\omega_1 = d \left(u(\rho) (\sigma_3 - \tilde{\sigma}_3)\right) =
\frac{\p u}{\p \rho} d\rho \w (\sigma_3 - \tilde{\sigma}_3) + u(
\sigma_1 \w \sigma_2 - \tilde\sigma_1 \w \tilde\sigma_2) \, , \eeq
where $u$ is an arbitrary function of $\rho$ which approaches a
constant $u_0$ at $\rho = \rho_0$ and dies away faster than
$1/\rho$ at $\rho \rightarrow \infty$. $\omega_1$ is harmonic for
the special choice
 \beq u = \frac{1}{\rho^2(\rho^2 +
6a^2)}\, .\eeq

In order to measure the correct RR two-form flux at infinity, we
need to fix the non-normalizable component of the flux. The
condition is that the total flux through both $\mathP^1$'s is $N$
units: \beq F|_{\rho \rightarrow \infty} = \frac{N}{2} \left(
\sigma_1 \sigma_2 + \tilde{\sigma}_1 \tilde{\sigma_2} \right) +
\ldots \, .\eeq The different ways of partitioning this flux
between the two $\mathP^1$'s cannot be distinguished at infinity
since we can change this partitioning by adding a suitable
multiple of $\omega_1$ to $F$. All partitions into a pair of
integers $(N_+, N_-)$ appear to give possible semi-classical
branches of moduli space: \beq F|_{\rho = \rho_0} = N_+ \sigma_1
\sigma_2 + N_- \tilde{\sigma}_1 \tilde{\sigma}_2 \, .\eeq In fact,
we should be careful here. A close consideration of the way in
which the orbifold group acts on the total space of the $U(1)$
fibre bundle reveals that the cases with $N_-$ even or odd can be
distinguished at infinity. We shall postpone a discussion of this
subtlety until the next section in which we lift to M-theory.

After fixing a choice of partition, a flux superpotential is
generated for the normalizable K\"ahler parameter. At large
volume, this flux superpotential is: \beq g_s W = \int F \w \omega
\w \omega = \int F \w (\omega_0 + \chi \omega_1) \w (\omega_0 +
\chi \omega_1) \, .\eeq $F$ is quantized and thus provides a fixed
background whilst $\chi$ is a fluctuating field. The vacuum
equations are: \beq \label{dW} \p_\chi W = \frac{2}{g_s} \int F \w
(\omega_0 + \chi \omega_1) \w \omega_1 = 0 \, . \eeq Using the
expression (\ref{omega1}) for $\omega_1$  we can integrate
(\ref{dW}) using Stokes' theorem to find boundary contributions at
$\rho=\rho_0$ and $\infty$. The contribution at $\rho = \infty$
vanishes whilst the contribution at $\rho = \rho_0$ vanishes if
$\chi$ is fixed so that: \beq \omega|_{\rho =\rho_0} \propto (N_+
\sigma_1 \sigma_2 - N_- \tilde{\sigma}_1 \tilde{\sigma}_2) \, .
\eeq  Thus the vacuum equations imply that the relative sizes of
the two $\mathP^1$'s are directly proportional to the fluxes
through them. We will find confirmation of this result when we
lift to M-theory in the following section.

In summary, we have found that in addition to the deformed
conifold branches, there is a collection of $O(K) \rightarrow
\mathP^1 \times \mathP^1$ branches labelled by two integers $(N_+,
N_-)$ corresponding to the fluxes through  the two $\mathP^1$'s.
Each of these branches is a (complex) one-dimensional family of
backgrounds, parametrized locally by the value of the
non-normalizable K\"ahler modulus. Furthermore, each background
has a unique vacuum state since the normalizable parameter is
frozen. There is a massless $U(1)$ field in each of the $O(K)
\rightarrow \mathP^1 \times \mathP^1$ backgrounds, since no mass
term is generated for the vector field in the same $\n=2$
multiplet as $\chi$.

\section{M-theory lift}

Next we lift the various families of solutions to M-theory. We
shall find that the orbifolds of the deformed and resolved
conifolds lift to orbifolds of the $G_2$ manifold $\mathR^4\times
S^3$ described in the introduction whilst the solutions with local
$\mathP^1\times\mathP^1$ lift to a new class of $G_2$ manifolds
found in \cite{Cvetic:2001kp}. These are all the solutions of
$G_2$ holonomy which asymptote to a unique $G_2$ cone over
$S^3/\mathZ_2\times S^3/\mathZ_N$ as we shall argue later using a
5-brane web analysis.

Each solution is labelled by the size of the minimal cycle in the
interior together with a 3-form period integral, and we recover
the unique $G_2$ cone in the limit of vanishing cycle. Therefore,
the classical moduli space consists of several (complex)
one-dimensional branches all meeting at the singular cone. In
discussing the modification of this picture due to quantum
effects, the low energy gauge symmetry is an important
characteristic of branches. The orbifolds of $\mathR^4\times S^3$
support no gauge symmetry, but the new solutions are expected to
have one normalizable harmonic 2-form and support a $U(1)$ gauge
symmetry. We shall use the number of $U(1)$ factors, $g=0$ or $1$,
as a label of branches.

\subsection{$g=0$ branch}

The configuration of $N$ D6-branes wrapped on the $\mathR\mathP^3$
of the $\mathZ_2$ orbifold of the deformed conifold lifts to an
orbifold of the familiar $G_2$ holonomy manifold of topology $\R^4
\times S^3$. In order to get $N$ D6-branes we orbifold the $\R^4$
by $\mathZ_N$. In addition we should orbifold by $\mathZ_2$ in
order to get an $S_3/\mathZ_2$. In the description in terms of
$(g_1,g_2,g_3)\sim (g_1h,g_2h,g_3h)$ with $g_1$ filled in to make
an $\mathR^4$, we have $\mathZ_2\times\mathZ_N$ acting as:
\begin{equation}
 (g_1, g_2, g_3) \sim (g_1, -g_2, g_3)
                 \sim (\omega g_1, g_2, g_3) ~,~~~\omega^N =1.
\end{equation}

The classical moduli space has two more $g=0$ branches
with $g_2$ or $g_3$ filled in.
In the second branch where $g_2$ is filled in, we have
$(\mathR^4/\mathZ_2)\times(S^3/\mathZ_N)$, and the M-theory circle
is identified with the Hopf fibre of $\mathZ_N$ lens space. The
IIA configuration is therefore a $\mathZ_2$ orbifold of the
resolved conifold with $N$ flux turned on. The $\mathZ_2$ reverses
the four directions transverse to the $\mathP^1$, and the
singularity supports an ${\cal N}=1$ $SU(2)$ SYM.

The classical geometry for the third branch with $g_3$ filled in
is the same as the second branch when $N$ is even, but is
different when $N$ is odd. To see this, notice that in the gauge
$g_2=1$ the $\mathZ_2$ acts on $g_1$ and $g_3$ as $-1$
simultaneously. If $N$ is even, one can redefine the generator of
$\mathZ_2$ by multiplying by the order two element of $\mathZ_N$
so that it acts only on $g_3$. Thus the second and the third
branches are identical in nature. On the other hand, for odd $N$
one cannot do this redefinition, and moreover one can see that the
$\mathZ_2$ acts freely on the third branch. The resulting IIA
geometry looks like the same $\mathZ_2$ orbifold of the resolved
conifold with $N$ flux, but this time the $\mathZ_2$ also acts on
the M-theory circle as a half-period shift. So the $\mathZ_2$
singularity does not support a gauge symmetry.

\paragraph{Quantum moduli space}

These three $g=0$ branches of classical moduli space are expected
to form a single smooth $g=0$ branch of the quantum moduli space.
Here we first briefly review  the arguments of
\cite{Atiyah:2001qf} for the case of $\mathR^4\times S^3$ without
orbifold, and then discuss the case of our interest according to
the analysis of \cite{Friedmann:2002ct}.

The branch structure of classical solutions $\mathR^4\times S^3$
is conveniently described by the $SU(2)$ matrices $g_{1,2,3}$
obeying the equivalence relation (\ref{equiv}). The three branches
are obtained by filling in one of the $g_i$ to make an $\mathR^4$.

Let $\hat{D}_j$ denote the $j$-th copy of $SU(2) \subset SU(2)^3$
and denote by $D_j$ the 3-cycle in $Y = SU(2)^3/SU(2)$ which is
the projection of $\hat{D}_j$ into $Y$. The $D_j$ obey an homology
relation
\[
 D_1+D_2+D_3=0.
\]
Furthermore, in the $j$-th branch, the cycle $D_j$ is trivial in
homology, and $Q_j=D_{j-1}=-D_{j+1}$ is non-vanishing. Define
three holomorphic parameters $\eta_{1,2,3}$ by
\[
\eta_i=V(D_i),~~~
 V(D)\equiv\exp \left(-\int_D(\Phi+iC)\right)
\]
using the harmonic 3-form $\Phi$ and the 3-form potential $C$.
They satisfy $\eta_1\eta_2\eta_3=-1$, where the minus sign comes
from a fermion anomaly \cite{Atiyah:2001qf}. Classically
$\eta_i=1, ~\eta_{i+1}\eta_{i-1}=-1$ on the $i$-th branch, but
quantum mechanically all the three classical branches are on a
single smooth Riemann sphere. Using the coordinate $z$ such that
the three large volume limits are corresponding to $z=0,1,\infty$,
one finds
\begin{equation}
  \eta_1=1-z,~~~
  \eta_2=\frac 1z,~~~
  \eta_3=\frac{z}{z-1}.
\end{equation}

Let us then take the $\mathZ_N\times\mathZ_2$ orbifold of this space,
where the $\mathZ_N$ and $\mathZ_2$ of the orbifold group act on $g_1$
and $g_2$ from the left.
By similar projections one finds the 3-cycles and the relation
\[
 D'_1=S^3/\mathZ_N,~~~
 D'_2=S^3/\mathZ_2,~~~
 D'_3=\left\{\begin{array}{lll}
      S^3/\mathZ_2,   & \frac N2D'_1+D'_2+ D'_3=0   & (N~{\rm even})\\
      S^3            ,& ND'_1+2D'_2+  D'_3=0   & (N~{\rm odd})
      \end{array}\right.
\]
The non-vanishing cycle in each branch is given by
\begin{equation}
 Q'_1=S^3/\mathZ_2,~~~
 Q'_2=S^3/\mathZ_N,~~~
 Q'_3=\left\{\begin{array}{lll}
      S^3/\mathZ_{N}   & (N~{\rm even})\\
      S^3/\mathZ_{2N}  & (N~{\rm odd})
      \end{array}\right.
\label{Q'}
\end{equation}
Classically in branches {\sf 1,2,3} the cycles obey some homology
relations as summarized in the table \ref{table:DQ}.
\vskip1mm
\begin{table}[htb]
\centerline{
\begin{tabular}{|c|c|c|c|}
\hline
$N$&
branch~{\sf1}~($D'_1=0$)&branch~{\sf2}~($D'_2=0$)&branch~{\sf3}~($D'_3=0$)\\
\hline
{\rm even}
 &$D'_2=-Q'_1,~D'_3=Q'_1$
 &$D'_3=-\frac{N}2Q'_2,~D'_1=Q'_2,$
 &$D'_1=-Q'_3,~D'_2=\frac{N}2Q'_3$\\
{\rm odd}
 &$D'_2=-Q'_1 ,~D'_3=2Q'_1$
 &$D'_3=-NQ'_2,~D'_1=Q'_2$
 &$D'_1=-2Q'_3,~D'_2=NQ'_3$\\
\hline
\end{tabular}}
\caption{homology relations among various cycles.}
\label{table:DQ}
\end{table}
The good local coordinates on the quantum moduli space around the
three large volume points are the fractional instanton factors.
For example on the first branch we have $SU(N)$ gauge symmetry and
so the fractional instanton factor is $z_1=V(Q'_1)^{1/N}$. We
define $z_2$ and $z_3$ similarly as the fractional instanton
factors on the other two branches.

\begin{figure}[htb]
\centerline{
\includegraphics{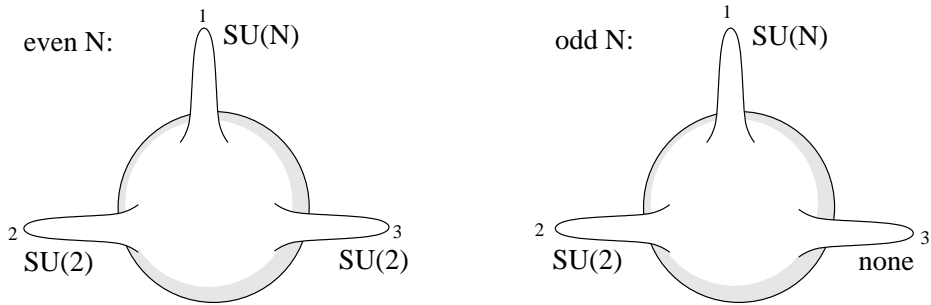}}
\caption{$g=0$ branch of quantum moduli space.}
\label{fig:gzero}
\end{figure}

Defining $\eta_i = V(D'_i)$ as before, one finds
$\eta_1^{N/2}\eta_2\eta_3=1$ for even $N$ and
$\eta_1^N\eta_2^2\eta_3=1$ for odd $N$. Note that the sign must be
determined from a careful analysis of the fermion anomaly,
although we fix it by requiring consistency with the parametric
representation of $\eta_i$ given below. Near the three
large-volume points the $\eta_i$ behave as follows; \vskip2mm
\begin{center}
\begin{tabular}{|c|c|c|c|}
\hline
$N$~even & $z_1\sim0$ & $z_2\sim0$ & $z_3\sim0$ \\
\hline
$\eta_1$ & $1$        & $z_2^2$    & $z_3^{-2}$ \\
$\eta_2$ & $z_1^{-N}$ & $1$        & $z_3^N$    \\
$\eta_3$ & $z_1^N$    & $z_2^{-N}$ & $1$        \\
\hline
\end{tabular}
~~~
\begin{tabular}{|c|c|c|c|}
\hline
$N$~odd  & $z_1\sim0$ & $z_2\sim0$ & $z_3\sim0$ \\
\hline
$\eta_1$ & $1$        & $z_2^2$    & $z_3^{-2}$ \\
$\eta_2$ & $z_1^{-N}$ & $1$        & $z_3^N$    \\
$\eta_3$ & $z_1^{2N}$ & $z_2^{-2N}$& $1$        \\
\hline
\end{tabular}
\end{center}
\vskip2mm
Assuming that the $g=0$ branch of the quantum moduli space is
topologically a sphere and introducing the coordinate $z$ as before,
one finds $\eta_i$ are expressed as the following functions of $z$:
\begin{equation}
 \eta_1=(1-z)^2,~~~
 \eta_2=z^{-N},~~~
 \eta_3=\left\{\begin{array}{ll}
      z^N/(1-z)^N       & (N~{\rm even})\\
      z^{2N}/(1-z)^{2N} & (N~{\rm odd}).
      \end{array}\right.
\label{eta-z}
\end{equation}
One is thus lead to the $g=0$ branch of quantum moduli space as in
 figure \ref{fig:gzero}.

In the above we claimed that the orbifold of $\mathR^4\times S^3$
supports $SU$ type gauge symmetries at various asymptotic regions
of moduli space, and in particular there is no $U(1)$ factor.
Indeed, such a $U(1)$ would arise from the M-theory 3-form
dimensionally reduced along a normalizable harmonic 2-form on the
$G_2$ space, but there is no such 2-form even after the asymptotic
behavior of the metric is modified so that there is an M-theory
circle of finite radius at infinity \cite{Brandhuber:2001yi}. This
supports our earlier claim that the correct choice of gauge theory
on the D-branes should be $SU(N)$ rather than $U(N)$.

\subsection{$g=1$ branches}

In this subsection we introduce a class of $G_2$ holonomy metrics
found in \cite{Cvetic:2001kp}. We claim that these solutions (with
a suitable orbifold action) represent the M-theory lift of the
IIA geometry with local $\mathP^1 \times \mathP^1$.
They take the form:
\begin{equation}
 ds^2 ~=~ dr^2 + a^2((\sigma_1+h\tilde\sigma_1)^2
                    +(\sigma_2+h\tilde\sigma_2)^2)
               + b^2(\tilde\sigma_1^2+\tilde\sigma_2^2)
               + c^2(\sigma_3-\tilde\sigma_3)^2
               + f^2(\sigma_3+h_3\tilde\sigma_3)^2
\end{equation}
where $a,b,c,f,h$ and $h_3$ are functions of the radial coordinate
$r$. The two sets of $SU(2)$ Maurer-Cartan forms $\sigma_a,
\tilde{\sigma}_a$ and the related $SU(2)$ matrices $X(\psi,
\theta, \phi),
\tilde{X}(\tilde{\psi},\tilde{\theta},\tilde{\phi})$ were defined
in section 2.

At large $r$, the metric asymptotes to the $S^1$ bundle over the
conifold
\begin{equation}
  ds^2 ~\stackrel{{\rm large}~r}\sim~ dr^2
 +\frac{r^2}{6}(\sigma_1^2+\sigma_2^2+\tilde\sigma_1^2+\tilde\sigma_2^2)
 +\frac{r^2}{9}(\sigma_3-\tilde\sigma_3)^2
 +f_\infty^2(\sigma_3+\tilde\sigma_3)^2.
\end{equation}
At small $r$ the coefficient functions behave as
\begin{equation}
\begin{array}{rcl}
 a&=&a_0+{\cal O}(r^2),\\
 b&=&b_0+{\cal O}(r^2),\\
 f&=&f_0+{\cal O}(r^2),
\end{array}
~~~
\begin{array}{rcl}
 c&=&  -r+{\cal O}(r^2),\\
 h&=&\frac{b_0f_0}{2a_0^3}r+{\cal O}(r^2),\\
 h_3&=&\frac{b_0^2}{a_0^2} +{\cal O}(r^2)\, .
\end{array}
\end{equation}
We see that a single $S^1$ shrinks at $r=0$ and in the IIA
geometry, upon reducing along the M-theory circle, we have a
$\mathP^1 \times \mathP^1$ bolt. Together with the asymptotic
behaviour at $r=\infty$ this ensures that the IIA reduction is
topologically $\mathO(K) \rightarrow \mathP^1 \times \mathP^1$.
In the limit that we shrink the M-theory circle to zero size,
the exact Calabi-Yau metrics (\ref{p1}) are recovered \cite{Cvetic:2001kp}.

We need to introduce an orbifold action in order to avoid a
conical singularity at $r=0$. Let us write $a_0^2= \lambda n_+,
b_0^2= \lambda n_-$ for some $\lambda \in \R$ which we shall fix
presently, and observe:
\begin{eqnarray*}
 ds^2 &\stackrel{r\sim 0}\simeq&
 dr^2 + r^2(d\psi-d\tilde\psi+\cos\theta d\phi-\cos\tilde\theta
d\tilde\phi)^2
 \\ &&
      + \frac{f_0^2}{n_+^2}(n_+d\psi+n_-d\tilde\psi
           +n_+\cos\theta d\phi+n_-\cos\tilde\theta d\tilde\phi)^2
 \\ && +ds^2_{S^2\times S^2}\mbox{(depends only on
               $r,\theta,\phi,\tilde\theta,\tilde\phi,\psi-\tilde\psi$)}
       ~+\cdots
\end{eqnarray*}
There will be a conical singularity unless the shrinking vector
\beq v = \frac{1}{n} \left( n_-\frac{\p}{\p \psi} -
n_+\frac{\p}{\p \tilde\psi}\right)\, ,\qquad n\equiv n_++n_- \, ,
\qquad |v|^2 = r^2 \eeq has period $2 \pi$. To remove this
singularity we must impose
\begin{equation}
  (\psi,\tilde\psi)
 \sim  (\psi,\tilde\psi) + \ts\frac{2\pi}{n}(n_-,-n_+).
 \end{equation}
Note that, in order for this identification to be a finite
orbifold action on $S^3\times S^3$, $n_+/n_- = a_0^2/b_0^2$ has to
be rational. We fix $\lambda$ by requiring that $(n_+,n_-)$ be a
pair of relatively prime positive integers. The bolt at $r=0$ is
the smooth space $T^{n_+,n_-}$ which is defined as
the set of $(X,\tilde{X})$ subject to the identification \beq (X,
\tilde{X}) \sim (X e^{i \alpha  n_- \tau_3} , \tilde{X} e^{-i
\alpha n_+ \tau_3}) \, . \eeq
An equivalent definition of $T^{n_+ \, n_-}$ is as the $U(1)$
bundle over $\mathP^1 \times \mathP^1$ with monopole numbers
$(n_+, n_-)$.
Note that the orbifold group contains the following element
\begin{equation}
 (\psi,\tilde\psi) \sim (\psi,\tilde\psi) +\ts\frac{4\pi}{n}(1,1).
\end{equation}
We may consider a further orbifold by \beq (\psi, \tilde\psi) \sim
(\psi , \tilde\psi) + \ts\frac{4 \pi}{N} (1,1)\, , ~~~~N\equiv k n
\, , (k \in \Z) \, , \eeq which does not lead to any new
singularities. With this choice of orbifold group, the total
D6-brane charge is $N = k n_+ + k n_- \equiv N_+ + N_-$. After
dimensional reduction on the M-theory circle to type IIA on
$\mathO(K) \rightarrow \mathP^1 \times \mathP^1$ we get
$(N_+,N_-)$ units of RR flux through the two $\mathP^1$'s. At
$r=0$ in the M-theory geometry, one has a non-vanishing $\mathZ_k$
orbifold of $T^{n_+,n_-}$ which is identified with $T^{N_+,N_-}$
under its definition above.

The restriction that $a_0^2/b_0^2$ be rational, which we have just
related to flux quantization in IIA, ensures that we have a
discrete set of one-parameter families of solution labelled by
$(\lambda;N_+,N_-)$, rather than a continuous two-parameter family
labelled by $(a_0,b_0)$. Also, note that the volumes of the
$\mathP^1$'s in the IIA geometry, which can be read off from the
$G_2$ metric at $r=0$, are proportional to the fluxes. This agrees
with the results of our IIA analysis using a flux superpotential.

Since the action of the orbifold group on $(X,\tilde X)$ is
\[
 (X,\tilde X) ~\sim~ (X\omega^{N_-/2},-\tilde X\omega^{N_-/2})
              ~\sim~ (X\omega,\tilde X\omega);
 ~~~~\omega = e^{\frac{2\pi i}{N}\tau_3},
\]
by using the relation $X=g_3g_1^{-1}, \tilde X=g_2g_1^{-1}$
one finds the action of orbifold on $(g_1,g_2,g_3)$
\begin{equation}
 (g_1,g_2,g_3) ~\sim~
 (\omega^{-N_-/2}g_1,-g_2,g_3) ~\sim~
 (\omega g_1,g_2,g_3);
 ~~~~\omega = e^{\frac{2\pi i}{N}\tau_3}.
\end{equation}
The structure of the orbifold group depends on whether $N_\pm$ are
even or odd.
\vskip2mm\noindent (i) If $N_\pm$ are both even and
therefore $N$ is even, the orbifold group is $\mathZ_2\times
\mathZ_N$. It contains a $\mathZ_2\times\mathZ_2$ subgroup, a
generator of which acts on $g_2$ as a sign flip and the other
flips $g_3$. So the orbifold group is invariant under the
permutation of $g_2$ and $g_3$.
\vskip2mm\noindent (ii) If $(N_+,N_-)=$ (odd,\ even),
the orbifold group is $\mathZ_2\times\mathZ_N$ with $\mathZ_2$
acting on $g_2$. Note that the orbifold group acts $g_2,g_3$
in an asymmetric way. If $(N_+,N_-)=$ (even,\ odd),
then the orbifold group is the same but now $\mathZ_2$ acts
on $g_3$. \vskip2mm\noindent (iii) If $N_\pm$ are both odd, the
orbifold group is $\mathZ_{2N}$ generated by $(\omega^{1/2},-1,1)$
or
\[
 (\omega^{1/2},-1,1)^{N+1}~\simeq~ (\omega^{1/2},1,-1).
\]
The orbifold group action is therefore symmetric under the
permutation of $g_2,g_3$.

\vskip2mm

Note the subtle dependence of the orbifold group on whether
$N_\pm$ is even or odd. At fixed $N$, the solutions asymptote to
two distinct orbifolds of the $S^1$ bundle over the conifold
depending on $N_-$ modulo 2. The solutions with even $N_-$ and
those with odd $N_-$ are therefore realizing two distinct classes
of four-dimensional ${\cal N}=1$ theories. This ties in nicely
with our assumption that there is no overall $U(1)$ gauge
symmetry. Recall that near the semiclassical point 1 of the
quantum $g=0$ branch drawn in figure \ref{fig:gzero} we expect
$SU(N)$ gauge symmetry and not $U(N)$. If the $\mathZ_2$ Wilson
lines were taking values in $U(N)$, the eigenvalues would be
$\pm1$'s with no further restriction. Since instead they take
values in $SU(N)$ the number of negative eigenvalues must be even.

Let us hereafter restrict to solutions with  even $N_-$. The
orbifold group acting on the asymptotic geometry is then the same
for all the allowed distributions $N=N_++N_-$, so we have to take
all of them into account in the discussion of the moduli space.
They all constitute $g=1$ branches of moduli space characterized
by the existence of an infrared $U(1)$ gauge symmetry.

\subsection{$T^3$ reduction and a type IIB dual}

We wish to explain what can be learnt about the moduli space of $G_2$
compactifications by making a Kaluza-Klein reduction on $T^3$. The
resulting base manifold is $\R^4$, as we explain below, and all
the topological information is contained in the details of how
cycles of the three-torus shrink.

First, we consider the case of the $G_2$ geometry of topology
$\R^4 \times S^3$ with no orbifold. Consider the $U(1)^3$ symmetry
group generated by:
\begin{equation}
\begin{array}{rcll}
(1,0,0) &:& (\- g_1, \+ g_2, \+ g_3),&
            (\+ X\+, \+ \tilde X\+), \\
(0,1,0) &:& (\+ g_1, \- g_2, \+ g_3),&
            (\+ X\-, \- \tilde X\-), \\
(0,0,1) &:& (\+ g_1, \+ g_2, \- g_3),&
            (\- X\-, \+ \tilde X\-).
\end{array}
\label{U1basis}
\end{equation}
Note that each $U(1)$ is normalized to have length $2 \pi$,
since \beq (- g_1, -g_2, -g_3) \sim (g_1, g_2, g_3) \, .\eeq

Now consider the Kaluza-Klein reduction on this $T^3$. A rough
argument that this leads to an $\R^4$ base is as follows. At $r
\neq 0$ the $U(1)^3$ action leaves fixed the coordinates $(r,
\theta, \tilde{\theta}, \psi - \tilde{\psi})$ whilst acting
transitively on the others. $\theta$ and $\tilde{\theta}$ each run
from 0 to $\pi$ and so together form a disc. On the boundaries of
the disc, ($\{\theta=0,\pi\} \cup \{\tilde{\theta} =0,\pi\}$), the
$\psi - \tilde{\psi}$ direction can also be gauged away and so
there is an $S^1$ fibred over the disc that shrinks at the
boundaries. This gives an $S^3$. At $r=0$, additional directions
shrink. In Phase 1, for example, the only non-shrinking direction
at $r=0$ is the $\tilde{\theta}$ line interval. Concentric $S^3$'s
shrinking onto a line interval at $r=0$ will lead to an $\R^4$
base.

Later, when we consider the $G_2$ manifold with local $T^{N_+,N_-}$,
the discussion is much the same except that the space at
$r=0$ is a two-disc spanned by $(\theta, \tilde{\theta})$. The
base is a family of concentric $S^3$'s flattening onto a two-disc
leading once more to $\R^4$.

Now we need to find which $S^1$'s shrink at special loci in the
geometry. It is straightforward to see that only the following
$S^1$'s can shrink at generic radii:
\begin{equation}
\begin{array}{rcl}
(1,0,0)&\leftrightarrow&(\theta=\pi,\tilde\theta=\pi), \\
(0,1,0)&\leftrightarrow&(\theta=0  ,\tilde\theta=\pi), \\
(0,0,1)&\leftrightarrow&(\theta=\pi,\tilde\theta=0  ), \\
(1,1,1)&\leftrightarrow&(\theta=0  ,\tilde\theta=0  ).
\end{array}
\end{equation}
To see that these cycles shrink, note that $\tau_3$ commutes with
X for $\theta=0$ and anticommutes for $\theta=\pi$.
Additionally, at $r=0$, an $S^3$ shrinks and so additional $S^1$'s
may vanish there.
In the $i$-th classical branch $g_i$ shrinks and so
\begin{equation}
\begin{array}{rcl}
(0,1,1)&\leftrightarrow& r=0~\mbox{(branch 1)}, \\
(1,0,1)&\leftrightarrow& r=0~\mbox{(branch 2)}, \\
(1,1,0)&\leftrightarrow& r=0~\mbox{(branch 3)}.
\end{array}
\end{equation}

We can now draw a diagram which represents this information. In
each case, the degeneration locus is three dimensional and two of
these dimensions lie within the $T^3$ fibre. Thus we can project
onto the $\R^4$ base and draw the degeneration loci as lines. The
degenerations which happen at arbitrary radius give semi-infinite
lines, whilst the others give line segments. The way in which the
lines end on each other can easily be read from the results above.
\begin{figure}[htb]
\centerline{
\includegraphics{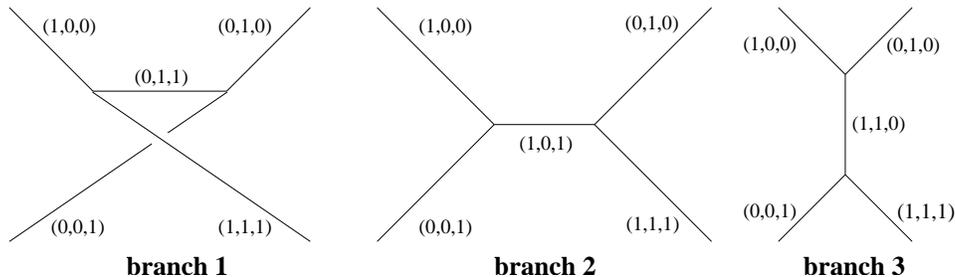}}
\caption{The three branches of the $\R^4 \times S^3$ geometry}
\label{fig:3dweb1}
\end{figure}

These graphs represent the three-branch structure of the classical
moduli space of M-theory on $\mathR^4\times S^3$. Each branch is
parametrized by the the volume of the minimal $S^3$ which is shown
as the length of the internal leg.

As was discussed in \cite{Leung:1997tw,Aganagic:2001jm}, by a
chain of dualities we can relate these diagrams with
three-dimensional webs of $(p,q,r)$ 5-branes in type IIB
compactified on $S^1\times S^1\times\mathR^4$, where $p,q$ denote
the NS5-brane and D5-brane charges and $r$ is the KK monopole
charge along the first $S^1$.

~

In the case of a $\mathZ_N$ orbifold acting on $g_i$ as $(g_1,
g_2, g_3)\to(\omega g_1, g_2, g_3),~\omega=e^{2\pi i\tau_3/N}$, we
should choose a different basis for our $T^3$. A suitable basis of
1-cycles of length $2 \pi$ is:
\begin{equation}
\begin{array}{rcl}
(1,0,0)' &:& (\- g_1, \+ g_2, \+ g_3),\\
(0,1,0)' &:& (\+ g_1, \- g_2, \+ g_3),\\
(0,0,1)' &:& (e^{\frac{i \alpha \tau_3}{N}}g_1,g_2,g_3).
\end{array}
\end{equation}

The cycles that shrink at arbitrary radius are the same as before
except that we need to take into account the change of basis. This
amounts to rewriting in terms of the new basis the labels of
four semi-infinite legs
\begin{equation}
(1,0,0)=(1,0,0)',~~~
(0,1,0)=(0,1,0)', ~~~
(0,0,1)=(0,-1,N)', ~~~
(1,1,1)=(1,0,N)',
\end{equation}
and those of the finite leg in each branch
\begin{equation}
(0,1,1)=N(0,0,1)',~~~
(1,0,1)=(1,-1,N)', ~~~
(1,1,0)=(1,1,0)'.
\end{equation}
Interestingly, in the first branch the orbifold group has fixed
points and this is reflected in the fact that the shrinking $S^1$
at $r=0$ has length $2 \pi /N$.
Correspondingly there will be $N$ coincident $(0,0,1)$ 5-branes at this
locus in the $(p,q,r)$ 5-brane picture.

~

Let us now turn to the cases of our interest.
First consider the geometries of $g=0$ given by
$\Z_N \times \Z_2$ orbifold of $\R^4 \times S^3$:
\begin{equation}
 (g_1, g_2, g_3)
 \sim (g_1, -g_2, g_3)
 \sim (\omega g_1, g_2, g_3),~~~
 \omega = e^{2 \pi i \tau_3/N}.
\end{equation}
A suitable basis of cycles of length $2 \pi$ for the $T^3$ is:
\begin{equation}
\begin{array}{rcl}
(1,0,0)'' &:& (e^{\frac{i \alpha \tau_3}{N}}g_1, g_2,g_3),\\
(0,1,0)'' &:& (g_1,e^{\frac{i \alpha \tau_3}{2}}g_2, g_3),\\
(0,0,1)'' &:& (\+ g_1,\+ g_2,\- g_3).
\end{array}
\end{equation}
The cycles which vanish at arbitrary radius are the same as in the
previous examples (this statement is true on any branch of classical
moduli space since it only depends on the asymptotic geometry). Moving to
the new basis we make the replacements of the labels of external legs:
\begin{equation}
(1,0,0)=(0,2,-1)'',~
(0,1,0)=(N,0,-1)'',~
(0,0,1)=(0,0,1 )'',~
(1,1,1)=(N,2,-1)''
\end{equation}
and those of the internal legs
\begin{equation}
(0,1,1)=N(1,0,0)'',~~~
(1,0,1)=2(0,1,0)'', ~~~
(1,1,0)=\left\{\begin{array}{ll}
  2(N/2,1,-1)'' & (N~{\rm even}), \\
  (N,2,-2)''    & (N~{\rm odd}).
 \end{array}\right.
\end{equation}
The data of the fibre degenerations is summarized in figure
\ref{fig:3dweb2} below.
\begin{figure}[htb]
\centerline{
\includegraphics{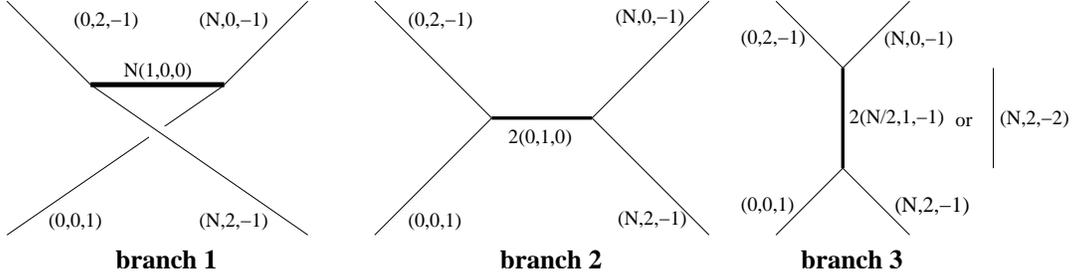}}
\caption{The three branches of $g=0$.}
\label{fig:3dweb2}
\end{figure}

\begin{figure}[htb]
\centerline{
\includegraphics{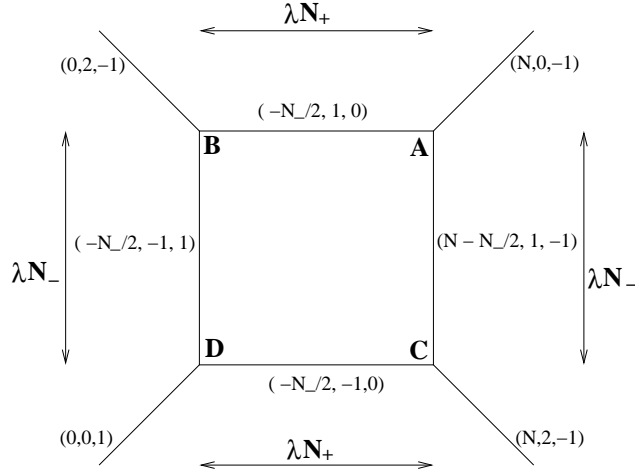}}
\caption{Classical geometries of $g=1$.}
\label{fig:3dweb3}
\end{figure}

Next consider the $g=1$ branches of the $G_2$ geometry.
In these branches, the following $S^1$ shrinks at $r=0$:
\begin{equation}
 (X, \tilde{X}) \rightarrow
 (X e^{\frac{i\alpha\tau_3N_-}{2N}},
  \tilde{X} e^{\frac{-i\alpha\tau_3N_+}{2N}}).
\label{shrinking}
\end{equation}
Once again we have normalized the generator so that the
$S^1$ has length $2 \pi$.
At generic points of the $r=0$ subspace, this $S^1$ does not lie
entirely inside the $T^3$. However, at special loci it does.
These are as follows:
\begin{equation}
\begin{array}{rclcl}
(N-\frac{N_-}{2},~~1,-1)'' &:&
(~~e^{\frac{i\alpha\tau_3}{2}}      Xe^{\frac{-i\alpha\tau_3N_+}{2N}},
 ~~~~~~~                     \tilde Xe^{\frac{-i\alpha\tau_3N_+}{2N}})
&\leftrightarrow& (r=0,\theta=0),\\
( -\frac{N_-}{2},-1,~~1)'' &:&
(e^{-\frac{i\alpha\tau_3}{2}}      Xe^{\frac{-i\alpha\tau_3N_+}{2N}},
 ~~~~~~~                    \tilde Xe^{\frac{-i\alpha\tau_3N_+}{2N}})
&\leftrightarrow& (r=0,\theta=\pi),\\
(-\frac{N_-}{2},-1,~~0)'' &:&
(~~~~~~~~                             Xe^{\frac{ i\alpha\tau_3N_-}{2N}},
 ~ e^{\frac{-i\alpha\tau_3}{2}}\tilde Xe^{\frac{ i\alpha\tau_3N_-}{2N}})
&\leftrightarrow& (r=0,\tilde\theta=0),\\
(-\frac{N_-}{2},~~1,~~0)'' &:&
(~~~~~~~~                            Xe^{\frac{ i\alpha\tau_3N_-}{2N}},
 ~~e^{\frac{i\alpha\tau_3}{2}}\tilde Xe^{\frac{ i\alpha\tau_3N_-}{2N}})
&\leftrightarrow& (r=0,\tilde\theta=\pi).
\end{array}
\end{equation}
The diagram for these geometries is presented in figure
\ref{fig:3dweb3}. The lengths of the internal legs, marked as
$\lambda N_+$, $\lambda N_-$ on the diagram, give the sizes of
three-cycles in the geometry.

It is an interesting exercise to read off various topological
properties of $T^{N_+,N_-}$ from this diagram. For example, the
$T^2$ fibration over internal legs {\bf BA} and {\bf DC} are both
lens space $S^3/\mathZ_{N_+}$ and are homotopic to each other, and
in a similar way the legs {\bf BD} and {\bf AC} both give
$S^3/\mathZ_{N_-}$. One can identify them with the lens spaces
within $T^{N_+,N_-}$ defined as submanifolds of fixed
$(\theta,\phi)$ or $(\tilde\theta,\tilde\phi)$. One can also read
off that the lens spaces $S^3/\mathZ_{N_\pm}$ obey one homology
relation so that the $G_2$ geometry has only one parameter.
Viewing the graph as a three-dimensional 5-brane web, one can also
check that the web is rigid except for overall rescaling. Later we
shall return to this 5-brane web picture to prove that there are
no further branches of classical moduli space.

\section{A dual type IIA picture}

We have described several families of $G_2$ metrics with the same
asymptotics and understood the structure of the classical moduli
space. We would now like to know how the corresponding quantum
moduli space looks. One nice way to study this problem will be to
try dimensional reduction along the diagonal $U(1)$ defined by
\[
 (g_1,g_2,g_3)~\to~
 (\omega g_1,\omega g_2,\omega g_3),~~~
 (X,\tilde X)~\to~
 (\omega X\omega^{-1},
  \omega \tilde X\omega^{-1})
~~~(\omega=e^{i\alpha\tau_3})
\]
This leads to a dual type IIA picture involving a special lagrangian
D6-brane in a non-compact Calabi-Yau manifold which admits a
gauged linear sigma model (GLSM) description.
The chief advantage of this description is that by moving to
the mirror IIB geometry we get an exact description of the
quantum moduli space through a certain curve contained in the
mirror target space\cite{Aganagic:2001jm}.

We begin in this section by translating the classical moduli space
obtained in the M-theory analysis to this framework and then study
the quantum moduli space within the mirror IIB picture in the
following section. The GLSM offers a quantitative description of
various blow-ups of the $\mathZ_N\times\mathZ_2$ orbifold in terms
of Fayet-Iliopoulos (FI) parameters. The Lagrangian D6-brane is
described as a half line in the toric base ending on its boundary.
The classical moduli space is therefore described by the FI
parameters and the position of the endpoint of the D6-brane.
However, as we will explain, some of the FI parameters are
effectively frozen due to the presence of the D6-brane.

\subsection{Toric description of the orbifold}

The three families of $\mathZ_N\times \mathZ_2$ orbifolds of
$\mathR^4\times S^3$ are mapped by the IIA reduction to
a Lagrangian D6-brane in the orbifold $\mathC^3/(\mathZ_N\times\mathZ_2)$
\begin{equation}
 (z_1,z_2,z_3)
~\sim~(z_1,e^{\frac{2\pi i}{N}}z_2,e^{-\frac{2\pi i}{N}}z_3)
~\sim~(-z_1,z_2,-z_3).
\end{equation}
The IIA reduction of the $G_2$ geometries with local $T^{N_+,N_-}$
should correspond to a certain blow-up of this orbifold with a
D6-brane at a suitable place.
Below we will describe these branches of classical moduli space
using toric geometry or GLSM.

\begin{figure}[htb]
\centerline{
\includegraphics{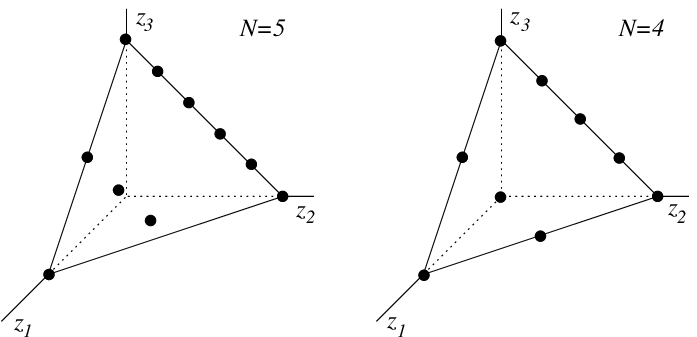}\hskip1cm
\includegraphics{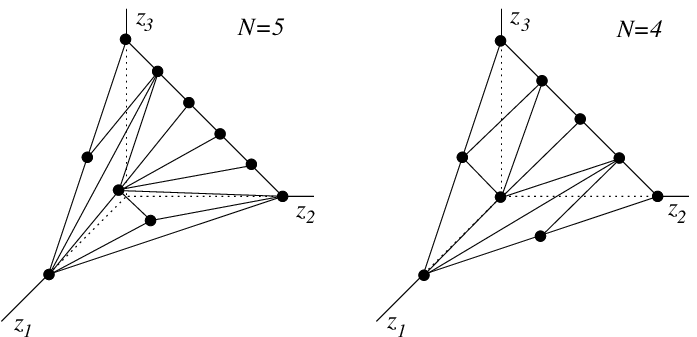}}
\caption{
The toric fans for the orbifolds $\mathC^3/(\mathZ_N\times\mathZ_2)$
and their maximal resolutions.}
\label{fig:toric1}
\end{figure}

We first find the toric data for the orbifold
$\mathC^3/(\mathZ_N\times\mathZ_2)$.
Introduce the coordinates $(z_1,z_2,z_3)$ on $\mathC^3$,
and associate to them the basis vectors $(1,0,0),(0,1,0),(0,0,1)$
of a three-dimensional lattice $\bf N$.
In toric geometry, they are the edge vectors and constitute
the toric fan of $\mathC^3$ made of a single cone
(positive octant of $\mathR^3$) and $\bf N$ is the lattice of
$\mathC^\times$ actions on $\mathC^3$:
\begin{equation}
 (n_1,n_2,n_3)\in{\bf N}:~~
 (z_1,z_2,z_3)\to (t^{n_1}z_1,t^{n_2}z_2,t^{n_3}z_3).
\end{equation}
The orbifold is described by the cone generated by the same
vectors, but the lattice $\bf N'$ of $\mathC^\times$ actions is
finer due to the orbifolding and is generated by
\[
 \rho_1=\ts\frac12(1,0,-1),~~\rho_2=\ts\frac1N(0,1,-1),~~\rho_3=(0,0,1).
\]
The toric fans are given in the two diagrams on the left of figure
\ref{fig:toric1} in the cases $N=5$ and $N=4$.

Let us refer to the triangle spanned by $(1,0,0),(0,1,0),(0,0,1)$
as $\Delta$ in what follows. The various toric blow-ups of the
orbifold singularity are described by the introduction of new edge
vectors and the subdivision of the positive octant into smaller
cones. For Calabi-Yau blow-ups, the new edge vectors should be
chosen from the lattice points on $\Delta$ which are depicted by
$\bullet$'s in the figure. One can include as many new edges as
one wishes, but the orbifold singularity is completely resolved
when all the possible edges are included. The two figures on the
right of figure \ref{fig:toric1} show examples of maximal
blow-ups.

The blown-up geometries are conveniently described by a GLSM as
classical moduli spaces of vacua. The GLSM associates a matter
field $z_i$ to each edge vector $v_i$, and a $U(1)$ gauge symmetry
$U(1)_a:~z_i\to e^{i\alpha Q_i^\alpha}z_i$ to each linear relation
among the edges, $\sum_iQ_i^av_i=0$. For maximal blow-ups there are
$(k+3)$ matter fields and $k$ $U(1)$'s with $k\equiv[\frac
{3N}{2}]$. Different blown-up manifolds are then described by
different level sets of D-term conditions
\begin{equation}
  \sum_i Q_i^a|z_i|^2 = r_a,
\end{equation}
modulo gauge equivalence. The various subdivisions of the positive
octant of the previous paragraph (triangulations of $\Delta$)
describe the fact that $\{z_i,z_j,z_k\}$ can simultaneously vanish
if and only if $\{v_i,v_j,v_k\}$ form a cone. For a suitably
chosen basis of $U(1)$'s, the manifold has large blown up cycles
when the $r_a$'s are large and positive whilst the resolutions are
turned off for sufficiently negative $r_a$'s.

Regarding toric Calabi-Yau threefolds as $T^3$ or
$T^2\times\mathR$ fibrations, one can draw the toric skeletons
\cite{Leung:1997tw,Aganagic:2000gs} or webs \cite{Aganagic:2003db}
which describe where and how the fiber degenerates in the base.
These diagrams are convenient for describing the various branches
of classical moduli space.

\begin{figure}[htb]
\centerline{
\includegraphics{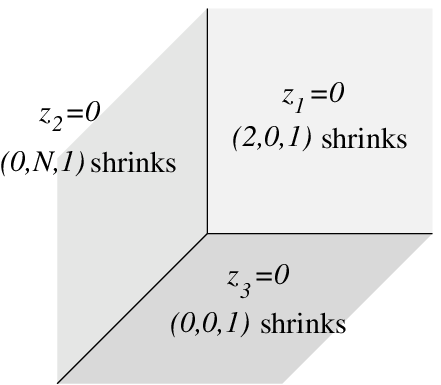}\hskip1cm
\includegraphics{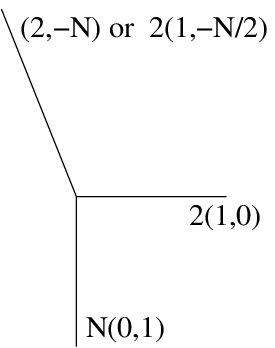}}
\caption{The skeleton and web for the orbifold.}
\label{fig:toric2}
\end{figure}

Let us first look at the orbifold with no resolution modes turned
on, corresponding to the $g=0$ branch. The skeleton is obtained in
the following way. Choose as the basis 1-cycles of the $T^3$
fiber, the three $U(1)$'s corresponding to the lattice points
$\rho_{1,2,3}$ and denote them by $\alpha_{1,2,3}$. Since
\begin{equation}
  (1,0,0) ~=~ 2\rho_1+\rho_3,~~~
  (0,1,0) ~=~ N\rho_2+\rho_3,~~~
  (0,0,1) ~=~ \rho_3,
\label{basisN}
\end{equation}
the three $U(1)$ moment maps $|z_i|^2$ generate translations along
the 1-cycles
\[
 |z_1|^2 ~\leftrightarrow~ 2\alpha_1+\alpha_3,~~~
 |z_2|^2 ~\leftrightarrow~ N\alpha_2+\alpha_3,~~~
 |z_3|^2 ~\leftrightarrow~ \alpha_3.
\]
The base of the $T^3$ fibration is identified with the first
octant of an $\mathR^3$ parametrized by the moment maps $|z_i|^2$.
The $T^3$ fibre  degenerates to a $T^2$ at generic points on the
boundary where one of the $z_i$ vanishes, and further to an $S^1$
on the three coordinate axes. This information is summarized in
the skeleton diagram given on the left of figure \ref{fig:toric2}.

On each leg of the skeleton one can find a vanishing cycle of the
form $n_1\alpha_1+n_2\alpha_2$. By projecting the skeleton onto a
2-plane so that the leg with vanishing cycle
$n_1\alpha_1+n_2\alpha_2$ is lying along the vector $(n_1,n_2)$,
we obtain a web diagram as shown on the right of figure
\ref{fig:toric2}.

A nice thing about the web is that it can be directly interpreted
as the $(p,q)$ 5-brane web in a IIB dual \cite{Leung:1997tw}. The
$(p,q)$-charges of the 5-brane legs and their angles at junction
points are simply understood from supersymmetry, charge
conservation and balance of the tension.

Let us also present the toric fan, skeleton and web diagrams for a
partial blow-up of the orbifold where only one mode corresponding
to the edge
\begin{equation}
 (\frac12, \frac{N_-}{2N},\frac{N_+}{2N})
 ~=~ \rho_1 + \frac{N_-}{2}\rho_2 + \rho_3
\end{equation}
is turned on. Blow-ups of this form will turn out to describe the
$g=1$ branches after the D6-brane is added. The GLSM now has four
fields $z_{1,..,4}$ obeying a D-term condition
\[
 N|z_1|^2 + N_-|z_2|^2 + N_+|z_3|^2 - 2N|z_4|^2  ~=~  r.
\]
and a $\mathZ_N$ orbifold identification
$(z_1,z_2,z_3)\sim(z_1,\omega z_2,\omega^{-1}z_3),~\omega^N=1$.
The orbifold singularity is partially blown up for positive $r$.
If $k$ be the greatest common divisor of $N_+, N_-$ and
$(N,N_+,N_-)=k(n,n_+,n_-)$, then the resolved target space is
a $\mathZ_k$ orbifold of the canonical bundle over
a weighted projective space $\mathW\mathP^2_{n,\;n_+,\;n_-}$.
The coordinates $z_{1,2,3}$ cannot vanish simultaneously so the
tip of the positive octant is chopped off from the base of $T^3$
fibration. See figure \ref{fig:toric3} below.
\begin{figure}[htb]
\centerline{
\includegraphics{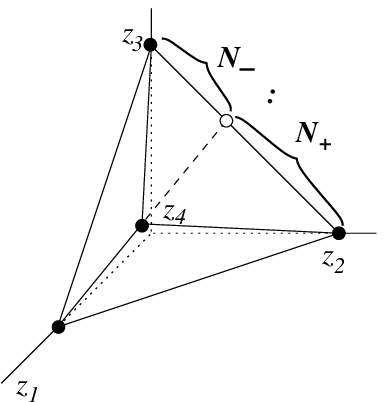}\hskip1cm
\includegraphics{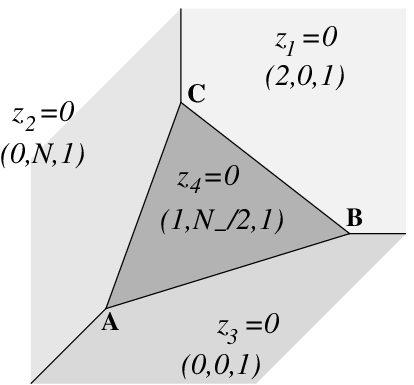}\hskip1cm
\includegraphics{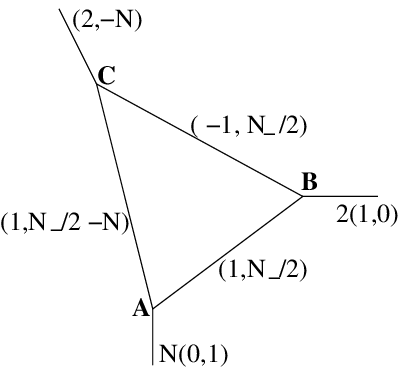}}
\caption{A partial resolution of the orbifold singularity.}
\label{fig:toric3}
\end{figure}

Similarly, turning on modes corresponding to vectors sitting on
the edges of the triangle $\Delta$ (i.e., lattice points on the
boundary of $\Delta$ excluding the three vertices) will chop off
the base of the $T^3$ fibration along the coordinate axes, and in
particular will increase the number of semi-infinite legs in the
skeleton or web diagrams. These are all resolutions of a line of
$\mathC^2/\mathZ_N$ or $\mathC^2/\mathZ_2$ orbifold singularities.
Although non-normalizable, one is free to turn on these modes to
deform the theory in the absence of the Lagrangian D6-brane. After
adding the D6-brane the situation changes drastically as we shall
explain later.

\begin{figure}[htb]
\centerline{
\includegraphics{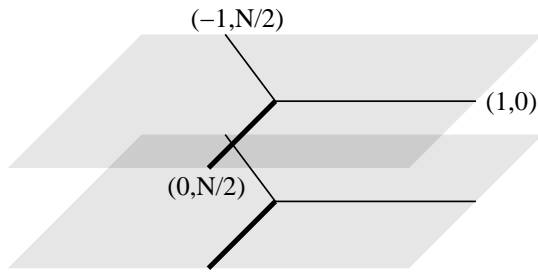}}
\caption{A deformation for even $N$.}
\label{fig:toric4}
\end{figure}

For even $N$, one can also deform the orbifold singularity so that
the web consists of two disjoint parts, each of which is made of
three legs with charges $(1,0), (0,N/2)$ and $(-1,N/2)$ as in
figure \ref{fig:toric4}. The deformation creates a 3-cycle of
topology $S^2\times S^1$, corresponding to a vertical line segment
stretching between parallel 5-brane legs.

\subsection{Lagrangian D6-branes}

Lagrangian D6-branes of topology $\mathC\times S^1$ are described
by half-lines in the toric base ending on a leg of the skeleton
\cite{Aganagic:2000gs, Aganagic:2001nx}. In the GLSM description
of a toric CY involving with $k+3$ coordinates, $z_i$, obeying $k$
D-term equations and a $U(1)^k$ equivalence relation, the relevant
branes have one direction along the toric base and two along the
fiber. They are therefore defined by two further linear
constraints on $|z_i|^2$ and one linear constaint on arg($z_i$).
The constraint on arg($z_i$) has to be
\[
 \sum_i{\rm arg}z_i={\rm const}
\]
for the submanifold to be {\it special}, and it follows that the
$k+2$ constraints on $|z_i|^2$ are all of the form
\[
 \sum_iq_i|z_i|^2={\rm const}, ~~~~(\sum_iq_i=0).
\]
The solution to the constraints in the base is:
$|z_i|^2=\xi+c_i,~(\xi\in\mathR_{\ge0})$. The topology of the
brane requires two of the $|z_i|^2$ vanish at $\xi=0$, which means
that the D6-brane must end on a leg of the skeleton.

Classical configurations are therefore described by a half-line
ending on a toric skeleton. The three $g=0$ branches are described
by a D6-brane ending on one of the three legs of the skeleton in
figure \ref{fig:toric2}, and the complex modulus corresponds to
the position of the endpoint and a $U(1)$ Wilson line on the
brane. For $g=1$ branches, we claim that the D6-brane has to end
at a certain point on the leg {\bf BC} of the skeleton in figure
\ref{fig:toric3}, and the only parameter is the overall scale of
the graph. We will verify this by moving to the web picture.

In the web diagram, D6-branes are described as half lines ending
on one of the legs and extending orthogonally to the web. However,
after moving to the 5-brane web picture in type IIB they turn into
TN5-branes carrying KK monopole charges along an $S^1$ of the
target space\cite{Aganagic:2001jm} and bend the web, to produce
three-dimensional web of $(p,q,r)$ 5-branes. The three-dimensional
web of 5-branes derived in this way should agree with the one
obtained in the previous section by $T^3$ reduction of the
M-theory solutions.

Let us start with the geometries with $g=0$ described by the
two-dimensional 5-brane web with three legs
\[
 N(0,1),~~~2(1,0),~~~(2,-N)
\]
Let us then add a TN5-brane to make the web three-dimensional with
four semi-infinite legs. The new external leg has charge
$(0,0,1)$, and the other legs also acquire TN5-brane charges in
order to satisfy the charge conservation at junction points. These
charges are determined from the following two requirements. First,
each external leg should be made of a single brane and not of
several coincident branes, since our M-theory solutions did not
have any orbifold fixed loci extending toward infinity. Secondly,
when the four legs are connected together by one finite leg, the
following gauge symmetries should be realized on the internal leg:
\[
 {\sf 1}:~SU(N),~~~
 {\sf 2}:~SU(2),~~~
 {\sf 3}:~SU(2)\mbox{ or none }(N\mbox{ even or odd})
\]
One then finds that, up to redefinition of basis, the four legs
must have the charge vectors
\begin{equation}
 (0,N,-1),~~~(2,0,1),~~~(2,-N,1),~~~(0,0,1)
\label{extlegs}
\end{equation}
This agrees with the result of $T^3$ reduction shown in
figure \ref{fig:3dweb2}, up to trivial signs and permutations.

Next we consider the $g=1$ branches labelled by $(N_+,N_-)$ and
described by the two-dimensional web of figure \ref{fig:toric3}.
We first fix the charges of the four external legs as in
(\ref{extlegs}) and then try to find the charges of the internal
legs and the endpoint of the fourth leg. We obtain in this way the
same web as was drawn in figure \ref{fig:3dweb3}, and furthermore
find that the endpoint of the fourth leg should be at a point
${\bf D}$ on the leg {\bf BC} of figure \ref{fig:toric3}
satisfying ${\bf BD}:{\bf DC}=N_-:N_+$. This is necessary in order
for the loop inside the web to close once we require that the legs
lie in directions imposed by supersymmetry, and it nicely agrees
with results from the M-theory geometry and the earlier IIA flux
superpotential analysis.

~

Note that the blow-up modes corresponding to the edges on the
faces of $\Delta$ have disappeared, or have been effectively
frozen after introducing the D6-brane, since no external 5-brane
legs are made of coincident 5-branes. To recover and turn on those
modes, one first has to send the D6-branes off to infinity in an
appropriate manner and change the charge of the semi-infinite leg
-- otherwise the supersymmetry will be broken. Therefore, the
theories with non-zero such blow-up modes are infinitely far away
from the theories of our interest.

\subsection{A no-go result for $g\ge2$ branches}
We have identified the backgrounds with $g=0$, $1$ with 5-brane
webs of the same {\it genus}. This is not a coincidence, as the
mirror IIB geometry turns out to contain a Riemann surface of the
same genus and the genus is indeed related to the number of $U(1)$
gauge symmetries in the IIB setup.

It is worthwhile to look for possible web configurations of higher
genus. This should be considerably easier than finding new $G_2$
holonomy solutions with the required asymptotics. In fact, we
would like to prove the absence of such webs of higher genus.

Balancing tensions in the web requires that the $(p,q,r)$ charges
of the legs add up to zero at each junction point. In this sense
the $(p,q,r)$ are better regarded as currents. Supersymmetry
requires that each leg of the web has to lie along a direction
determined by its charge vector. Namely, a leg with $(p,q,r)$
5-brane charge has to extend along the $(p,q,r)$ direction. Let us
introduce a height function on the vertices of the web which gives
the position in the $(0,0,1)$ direction. We choose our convention
so that $r$-current always flows downhill.

\begin{figure}[htb]
\centerline{
\includegraphics{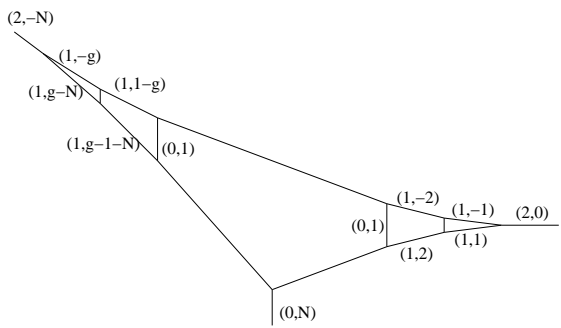}}
\caption{the two-dimensional web for maximal blow-up}
\label{fig:nogo1}
\end{figure}

From our previous discussion  we may restrict attention to webs
which have four legs with charges given by (\ref{extlegs}), and
which admit a projection to a two-dimensional web corresponding to
a blow-up of the orbifold $\mathC^3/(\mathZ_N\times\mathZ_2)$. We
can therefore only turn on blow-up modes corresponding to edges in
the interior of $\Delta$, and the genus is at most
$[\frac{N-1}{2}]$. A two-dimensional web of maximal genus $g$ is
shown in figure \ref{fig:nogo1}. The body of the web has the
structure of a large triangle partitioned by parallel finite legs
of $(p,q)=(0,1)$. What we would like to prove
 therefore is the following \vskip2mm\noindent {\bf No-go result}:~
{\it there is no supersymmetric 5-brane web of genus $\ge2$ with
four external
     legs of charges (\ref{extlegs}) which admits a projection
     to a two-dimensional web of the type drawn in figure \ref{fig:nogo1}.
}

\begin{figure}[htb]
\centerline{
\includegraphics{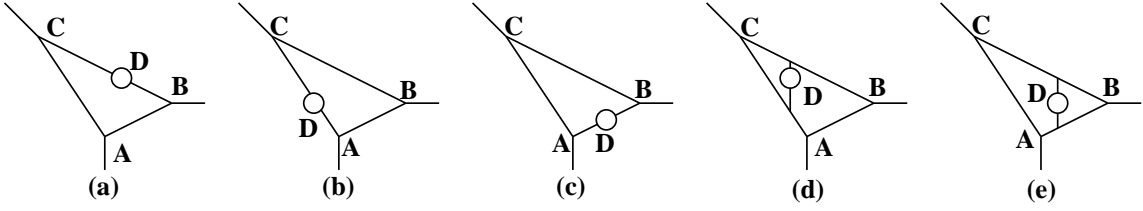}}
\caption{the five choices for the location of the $(0,0,1)$
5-brane.} \label{fig:nogo2}
\end{figure}

In the following we will denote by {\bf A,B,C,D} the vertices
where the four external legs of charges
$(0,N,-1),(-2,0,-1),(2,-N,1),(0,0,1)$ are attached to the body of
the web\footnote{We have reversed the sign of the (-2,0,-1) charge
vector relative to our previous convention. This is so that the
total charge flowing into the diagram is vanishing. For the
present argument we need to be careful of such sign choices.}.
There are five possible ways of attaching {\bf D} to the web as
shown in figure \ref{fig:nogo2}.

The external legs at {\bf A} and {\bf B} are sources of a unit
$r$-current whilst the legs at {\bf C} and {\bf D} are sinks. The
height function is therefore not constant on the the web. We begin
by trying to identify the lowest and highest vertices.

Since $r$-current flows downhill, there can be no current flowing
out of the lowest vertex. Thus the lowest vertex must either be
one of the sinks {\bf C}, {\bf D} or else a vertex at which there
is no flow of current in or out. In the latter case, all
neighbouring vertices will also be lowest vertices and this will
continue along any path until we hit ${\bf C}$ or ${\bf D}$.

A brief inspection of figures \ref{fig:nogo1} and \ref{fig:nogo2}
shows that every internal vertex on the web can be connected
either to ${\bf A}$ or to ${\bf B}$ by a path which avoids ${\bf
C}$ and ${\bf D}$. Thus if any internal vertex is lowest then, by
following the path, we would find that ${\bf A}$ or ${\bf B}$ is
also a lowest vertex which is a contradiction since they are
sources for current. We can conclude that ${\bf C}$ and ${\bf D}$
are the lowest vertices and also that they are neighbours with no
junction points in between.

By a similar reasoning, ${\bf A}$ and ${\bf B}$ are the highest
vertices and are also neighbours. These requirements prohibit webs
of types (c),(d),(e) in figure \ref{fig:nogo2}. The only allowed
webs are those of type (a) or (b) with no junction points on {\bf
AB} and {\bf CD}.

Now let us make a basis change of the charge vectors
$(p,q,r)=(p,q,r-p)'$. This will change the labels of external legs
as
\[
 {\bf A}(0,N,-1)',~~~~
 {\bf B}(-2,0,1)',~~~~
 {\bf C}(2,N,-1)',~~~~
 {\bf D}(0,0, 1)'
\]
and in particular the role of ${\bf B,C}$ as a source and a sink
of $r$-current are exchanged. This redefinition of charges
transforms the webs (b) and (c) of figure \ref{fig:nogo2} into
each other, so if one is inconsistent so is the other. Applying
this redifinition to the web (a) one finds that the paths of
highest and lowest points are now given by {\bf AC} and {\bf BD}.
There should be no junction points on these legs as well as {\bf
AB} and {\bf CD}, so the internal legs of $(p,q)=(0,1)$ cannot end
anywhere. We have shown that the only consistent webs are of type
(a) with no additional legs. They are precisely the $g=1$
configurations discussed in the previous subsection.

\section{IIB mirror}

Now we turn to the analysis of the quantum moduli space. We would
like to obtain its holomorphic structure by working in the mirror
type IIB picture. The target space is given by a hypersurface in
$\mathC^4$:
\begin{equation}
\xi\eta=F(u,v;t_i),
\label{mirrorCY}
\end{equation}
where $t_i$ parametrizes the complex structure. The curve
$\Sigma:F(u,v;t_i)=0$ has the same topology as the toric skeleton
of the previous section with the legs ``fattened''. The lagrangian
D6-brane turns into a D5-brane at $\xi=0$ extending along
$\eta$-direction and intersecting with the curve $\Sigma$ at a
point $(u_0,v_0)$. The superpotential as a function of $(u,t_i)$
is given by\cite{Aganagic:2000gs, Aganagic:2001nx}
\begin{equation}
  W(u,t_i) = \int_{u^*}^{u_0}v(u;t_i)du.
\label{W}
\end{equation}
with a suitably chosen reference point $(u^\ast,v^\ast)$. In the
following, we will first derive the curve $\Sigma$ starting from
the GLSM of the previous section, clarifying in particular the
relation between $t_i$ and the complexified K\"ahler moduli on the
IIA side. We will then give a precise form of the superpotential
and the F-term equations for our system. Finally the exact
holomorphic structure of the  quantum moduli space will be
obtained for some examples with small $N$.

For mathematical background on complex curves, we refer the reader
to \cite{Griffiths-Harris}.

\subsection{Curve within the mirror geometry}

Let us first derive the curve $\Sigma$ for our
$\mathZ_N\times\mathZ_2$ orbifold of $\mathC^3$.
We should start with the linear sigma model in the IIA side with
all the blow-up modes taken into account.
There are therefore $k+3$ chiral matter fields $z_i$
obeying $U(1)^k$ equivalence relations and $k$ D-term conditions,
with $k=[3N/2]$.
Let us regard $z_{1,2,3}$ as the original coordinates of
$\mathC^3/(\mathZ_N\times\mathZ_2)$ and the rest as corresponding to
blow-ups, and choose the set of FI parameters $r_a$ so that the D-term
conditions take the form
\begin{equation}
  \sum_{i=1,2,3}Q_i^a|z_i|^2~+~
  \sum_{b=1}^kQ_{3+b}^a(|z_{3+b}|^2+r_b) ~=~ 0.
\end{equation}
We will refer to $r_b$ as the blow-up mode corresponding to the
edge vector for $z_{3+b}$. The mode is turned on or off when $r_b$
is large positive or negative. Indeed, when $r_b$ is large
negative the field $z_{3+b}$ has to condense and become massive
along with one of the $U(1)$'s, so they disappear from the
low-energy physics.

Mirror symmetry \cite{Hori:2000kt} transforms these matter fields
into $k+3$ twisted chiral fields $Y_i$ related by ${\rm
Re}Y_i=|z_i|^2$, and maps the GLSM to a Landau-Ginzburg model.
These twisted matter fields obey $k$ linear relations that follow
from the D-term conditions. They are solved in terms of three
fields $u,v,w$ as
\[
 Y_i ~=~ a_iu + b_iv + c_iw - t_i;~~~~
 \left(\begin{array}{ll}
 t_{1,2,3}=0,\\
 t_{3+b}\;= r_b+i\theta_b&(b=1,\cdots,k)
 \end{array}\right)
\]
where $(a_i,b_i,c_i)$ is the edge vector
$v_i=a_i\rho_1+b_i\rho_2+c_i\rho_3$ corresponding to $z_i$, and
$r_b+i\theta_b$ are complexified K\"ahler parameters. The LG model
of $\{u,v,w\}$ has the superpotential $W_{\rm LG}=\sum_i
e^{-Y_i}$. Since $c_i=1$ for all the edge vectors of our toric fan
under the choice of lattice basis (\ref{basisN}), the dependence
of $W_{\rm LG}$ on $w$ factorizes:
\[
 W_{\rm LG} ~=~ e^{-w}\sum_i e^{-a_iu-b_iv+t_i}
   ~=~ e^{-w}F(u,v;t_i).
\]
As explained in \cite{Hori:2000ck}, this LG model is equivalent
to the sigma model on a CY space (\ref{mirrorCY}).

\begin{figure}[htb]
\centerline{\includegraphics{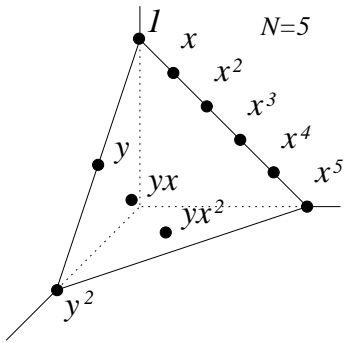}}
\caption{Monomials of $x,y$ corresponding to edges of the fan.}
\label{fig:fan-xy}
\end{figure}

Introducing $e^{-u}=y,e^{-v}=x$ one finds that each edge vector is
associated to a monomial $y^{a_i}x^{b_i}$ in $F$ as in figure
\ref{fig:fan-xy}. The curve $\Sigma$ is expressed in terms of
$x,y$ as
\begin{equation}
 F(x,y;t_i) ~=~
 y^2 - 2yP(x;t_i) + Q(x;t_i) ~=~0.
\label{curve1}
\end{equation}
where $P$ and $Q$ are polynomials of $x$ of degree $[\frac{N}{2}]$
and $N$ respectively, and the coefficient of each monomial is
given in terms of the corresponding K\"ahler parameter as
$e^{t_i}$. The curve therefore has genus $[\frac{N-1}{2}]$ for
generic values of the K\"ahler parameters.

Looking at the toric fan, there are as many edge vectors in the
interior of $\Delta$ as the genus $g$ of $\Sigma$. These
correspond to normalizable deformations of the target space.
Indeed, for each 1-cycle $\alpha$ of the curve $\Sigma$ there
corresponds a 3-cycle defined by a circle fibration over a 2-disc
bounded by $\alpha$, with the fiber being the nontrivial $S^1$ of
the cylinder $\xi\eta=F(x,y)$ in $\xi$-$\eta$ space. One can count
normalizable deformations of complex structure by counting
mutually non-intersecting compact 3-cycles, and there are $g$ of
them. Also, there arise $g$ $U(1)$ gauge fields from dimensional
reduction of the IIB 4-form potential along the dual 3-cocycles.

On the other hand, the edge vectors sitting on the faces of
$\Delta$ correspond to non-normalizable deformations as they alter
the asymptotic form of the curve. Recall that the variables $x,y$
are defined from the LG fields, so that they are
$\mathC^\times$-valued. The curve $\Sigma$ written in terms of
$x,y$ therefore has several punctures. The coefficients of
monomials on the face of figure \ref{fig:fan-xy} connecting
$(0,1,0)$ and $(0,0,1)$ determine the location of the punctures at
$y=0$, and similarly for the other two faces. They are also
related to the separation of the parallel semi-infinite legs of
the skeleton or web in the IIA picture. As such, they are clearly
non-normalizable.

Depending on whether $N$ is even or odd, the curve has $(N+2+2)$
or $(N+2+1)$ punctures for generic choices of the non-normalizable
complex structure parameters. The curve of interest to us should
have only three punctures, so we fix some of the coefficients in
the definition of curve by requiring that\cite{Aganagic:2001jm}
\begin{equation}
\begin{array}{ll}
{\sf 1}.&\mbox{$F(x,0)$ has a degenerate root of order $N$,} \\
{\sf 2}.&\mbox{$F(0,y)$ has a double root,} \\
{\sf 3}.&\mbox{$\hat F(y)\equiv{\rm lim}_{x\to\infty}x^{-N}F(x, yx^{N/2})$
               has a double root when $N$ is even.}
\end{array}
\end{equation}
One can rescale $x,y$ so that the first two punctures are
at $(x,y)=(1,0)$ and $(0,1)$.
The curve should thus take the form
\begin{equation}
 0=F(x,y)=y^2+(1-x)^N-2yP(x;s_i)
\label{curve2}
\end{equation}
with
\begin{equation}
\begin{array}{lrcl}
 (N=2n) &
 P(x;s_i) &=& 1+s_1x+\cdots s_nx^n,~~~~s_n\equiv\pm1\\
 (N=2n+1) &
 P(x;s_i) &=& 1+s_1x+\cdots s_nx^n
\end{array}
\label{s_i}
\end{equation}
We are left with normalizable deformations corresponding
to the interior points of $\Delta$.
In the following we denote by {\bf A},{\bf B},{\bf C} the three
punctures at $(x,y)=(1,0), (0,1), (\infty,\infty)$.

Note that, with no D5-branes and ${\cal N}=2$ spacetime
supersymmetry, there is nothing wrong with considering the whole
family of theories related by non-normalizable deformations. This
situation changes drastically when the D5-brane is added and a
superpotential (\ref{W}) is generated. Furthermore, we will see
that many of the normalizable deformations are fixed and the curve
cannot have genus $\ge2$ owing to the presence of the
superpotential.

\subsection{Superpotential}

Now we turn to the analysis of the superpotential (\ref{W})
generated by the D5-brane. The integral contains an undetermined
reference point $(u^*,v^*)$, so that $W$ has an ambiguity up to a
(possibly $t_i$-dependent) constant shift. We will first fix this
ambiguity by the following simple argument.

\begin{figure}[htb]
\centerline{
\includegraphics{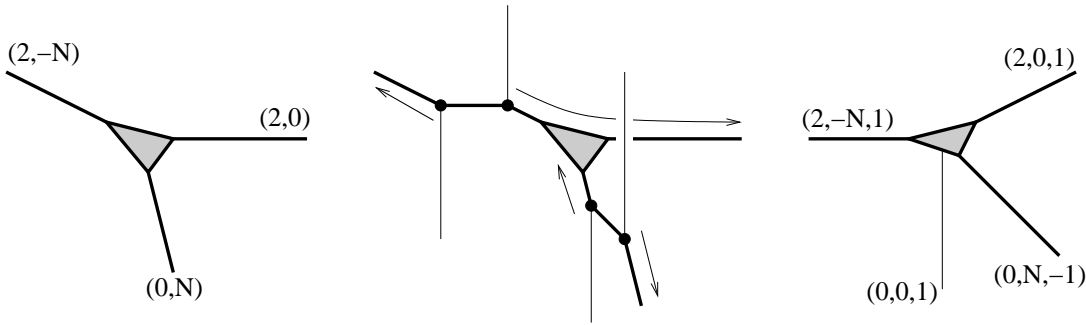}}
\caption{Construction of three-dimensional webs from two-dimensional ones.}
\label{fig:2d3d}
\end{figure}

Recall that in the previous section we obtained a description of
the moduli space in terms of three-dimensional webs with four
external legs
\[
 (0,N,-1),~(2,0,1),~(2,-N,1),~(0,0,1).
\]
Starting from a two-dimensional web with external legs $(0,N),
(2,0), (2,-N)$, one can reproduce the above set of legs as
illustrated in figure \ref{fig:2d3d}. We put two $(0,0,1)$ 5-brane
lines anywhere on the web. Once they intersect with the legs of
the web each of them breaks into halves. One sends three of the
four half-lines to infinity along the external legs of the
original web to alter their charges, thereby arriving at the
three-dimensional web with the required external legs.

The interpretation of this procedure in the mirror IIB side is
clear. A D5-brane with worldvolume $\xi\eta=F(u,v)=c$ decomposes
into a D5-brane with $\xi=0$ and a $\Dbar$5-brane with $\eta=0$
when it touches the curve, i.e. when $c$ vanishes. We should
therefore attach two pairs of D5, $\Dbar$5-branes to the curve,
and bring $\Dbar$5-branes to the punctures {\bf A},{\bf B} and a
D5-brane to {\bf C}. We will refer to the remaining free D5-brane
as {\bf D}. The superpotential is then given by an integral of the
form (\ref{W}) over the contour produced by the pairs of
D5-branes,
\begin{equation}
W = \int_{\bf B}^{\bf C}vdu + \int_{\bf A}^{\bf D}vdu.
\label{W2}
\end{equation}

Several comments are in order.
First, for curves with genus $g\ge1$, the integration
contour is still ambiguous due to the presence of non-trivial one-cycles.
This ambiguity is related to the fact that, by moving a D5-brane
 along a closed contour on the curve, one can induce a RR3-form flux.
So topologically different contours lead to different flux
backgrounds. One could also try writing the superpotential as
\[
W = \int_{\bf B}^{\bf D}vdu + \int_{\bf A}^{\bf C}vdu.
\]
This leads to the same superpotential as long as the difference
between the two contours is trivial;
\[
 ({\bf BC} +{\bf AD}) - ({\bf BD} +{\bf AC})
 = {\bf AD} + {\bf DB} + {\bf BC} + {\bf CA} ~=~ 0~\in H_1(\Sigma)
\]
We will present the solution to the F-term equations
$\partial_{s_j}W=0$ in a way that avoids complications due to this
ambiguity and the contour will be denoted simply as
$\int_{\Dbar5}^{\rm D5}$.

The second comment concerns the parameters of the theory. Moving
the free D5-brane {\bf D} is clearly a non-normalizable
deformation and in that sense would appear to be on an equal
footing with the deformations of the curve that change the
positions of punctures. However, as labels of theories one should
not regard these two kinds of parameters on the same footing. This
is because the superpotential (\ref{W2}) is not well-defined
without regularizing the divergence of the integral near each end
of the contour at the punctures. While the regulator is
independent of the position of {\bf D}, it should depend on the
position and multiplicity of punctures. As a result, the variation
of $W$ under the motion of {\bf D} still makes sense but the
variation under the non-normalizable deformation of the background
does not. Since we want the theories in a given family to be
defined by the same regularization, the position of {\bf D} is the
only relevant parameter of the theory\footnote{It may be helpful
to recall that the difference between the two types of parameters
is particularly clear in the
 5-brane web picture of the moduli space, where after the addition
 of the $TN$ 5-brane charge, the external legs of the web can no longer
 be separated.}.

Our superpotential has a further ambiguity up to $SL(2,\mathZ)$
linear transformations of $u,v$. Such a subtlety was discussed in
detail in \cite{Aganagic:2000gs, Aganagic:2001nx,
Aganagic:2001jm}. There it was shown that, under a suitable choice
of coordinates, the mirror IIB superpotential for a simpler
orbifold $(\mathC^2/\mathZ_N)\times\mathC$ exactly reproduces the
topological A-model amplitude for the relevant large $N$
transition of the conifold. We will not need to go into the
details of this ambiguity, since in our case it only shifts $W$ by
some constants independent of $s_i$, and we will only consider the
variations of $W$ with respect to the $s_i$.

~  Let us turn to the analysis of supersymmetric vacua of the
theory. We introduce new coordinates $x,\hat y$ to rewrite the
curve as
\begin{equation}
  \Sigma~:~
  y^2-2yP(x;s_i)+(1-x)^N ~=~
  \hat y^2 + (1-x)^N-P(x;s_i)^2 ~=~0
\label{curve3}
\end{equation}
The curve is then a standard double cover of the $x$-plane with
$2g+2$ branching points (including $x=\infty$). The superpotential
and its moduli-derivatives now read,
\begin{equation}
  W = \int_{\Dbar5}^{\rm D5}\frac{\log ydx}{x} ~~,~~~
  \frac{\partial W}{\partial s_j}
    = \int_{\Dbar5}^{\rm D5}\frac{x^{j-1}dx}{\hat y}
~~~~
  (j=1,\cdots,g=\ts[\frac{N-1}{2}]).
\end{equation}
The $s_j$-derivatives of $vdu$ give the set of $g$ independent
holomorphic 1-forms on $\Sigma$.

In mathematical terms, the formal difference between the set of
D5-branes and $\Dbar$5-branes defines a divisor $D$ of $\Sigma$.
Its is of degree zero, i.e., the number of D5-branes is the same
as the number of $\Dbar$5-branes. The F-term equations define a
map $\mu$ from divisors of degree zero to the Jacobian variety of
$\Sigma$
\[
 {\Jacobi}_g(\Sigma) = \mathC^g/\Lambda,
\]
where $\Lambda$ is the period lattice representing the ambiguity
in the choice of contours. The F-term conditions require $\mu(D)=0
\in\Jacobi_g(\Sigma)$, namely that the $g$ period integrals vanish
under a suitable choice of contours connecting D5-branes and
$\Dbar$5-branes.

By Abel's theorem, $\mu(D)$ vanishes modulo periods if and only if
there is a meromorphic function on $\Sigma$ with poles precisely
at D5-branes and zeroes at $\Dbar$5-branes. Therefore, in our
problem the F-term condition is solved by choosing the moduli
$s_i$ for which there exists a meromorphic function on $\Sigma$
with poles at $\bf C,D$ and zeroes at $\bf A,B$. Such a function,
if it exists, defines a map from $\Sigma$ to $\mathP^1$ of degree
two. It is known that on hyperelliptic curves there are always
such degree two maps $\Sigma\to\mathP^1$, but it is also known
that all such maps are related to the map $x$ by fractional linear
transformation. Since $\bf A,B,C$ are located at points with
different values of $x$, we conclude that hyperelliptic curves
$(g\ge2)$ cannot solve the F-term conditions. This is in perfect
agreement with the no-go result for webs of genus $\ge2$ which we
found earlier!

From the above argument it appears to follow that for $N\ge5$
there is no solution to the F-term condition, but this is not the
case. The important point is that the curve can have a smaller
genus for special choices of moduli due to degeneration of branch
points. For such degenerate curves of genus $g'\le g$, one can
consider reduced maps $\mu': {\rm
Div}^0(\Sigma)\to\Jacobi_{g'}(\Sigma)$. For $g'=1$ we expect a
solution to $\mu'(D)=0$ modulo periods. Of course, for $g'=0$ the
condition $\mu'(D)=0$ is vacuous.

What about the remaining $(g-g')$ conditions? To solve these, one
must recall that the drop in genus comes from a degeneration of
cycles in $\Sigma$. This is accompanied by the emergence of
massless monopole hypermultiplets associated to D3-branes wrapping
the corresponding shrinking 3-cycles of the CY. Let us be
schematic here and parametrize the moduli space of $\Sigma$ by
$(t_1,\cdots,t_g)$, and suppose that a monopole $M_i,\tilde M_i$
becomes light near the locus $t_i=0$~ ($i=1,\cdots,g-g'$). The
superpotential near such a locus is modified by terms containing
monopole fields,
\begin{equation}
  W(t) ~\longrightarrow~
  W(t) +\sum_{i=1}^{g-g'}t_iM_i\tilde M_i.
\end{equation}
The variation with respect to $M_i,\tilde M_i$ require $t_i=0$,
namely that the curve is of reduced genus. Moreover, the $(g-g')$
remaining conditions corresponding to the vanishing of the
$t_i$-variations, merely determine the values of the monopole
condensates.

We thus arrived at a rough understanding of the structure of the
quantum moduli space. We regard the position of the free D5-brane
$\bf D$ on $\Sigma$ as the coordinate on moduli space. For each
choice of the position of $\bf D$ there will be a certain number
of solutions to the F-term condition, which are curves of genus 0
or 1. The $g=1$ branches can meet with $g=0$ branches at points of
monopole condensation.

\subsection{Resolution of singularities}

The light monopoles are identified with D3-branes wrapped on certain
compact 3-cycles in the target space.
Each such 3-cycles is defined as a circle fibration over a 2-disc bounded
by the non-trivial 1-cycle $\alpha$ of $\Sigma$, so it shrinks
precisely when $\alpha$ shrinks.
The target space then develops a conifold singularity,
since near the degenerate branch point of $\Sigma$ at $x=x_0$
the target space is locally described by
\[
 \xi\eta - \hat y^2 + c(x-x_0)^2+{\cal O}((x-x_0)^3)=0.
\]
The condensation of D3-branes blows up the singularity as
discussed in \cite{Strominger:1995cz,Greene:1995hu}. The
double-point singularity of the curve is also resolved into two
distinct points on the blown-up $\mathP^1$. The 1-cycle $\beta$
which had non-zero intersection number with $\alpha$ is pinched
off, and we end up with a smooth curve of lower genus in a
blown-up target space.

One has to require that the degeneration of branch points of the
curve should not occur at the points {\bf A},{\bf B},{\bf C},
because this would change the asymptotics of the curve. We will
call curves with degenerate branch points at punctures {\it
singular}. A curve is singular when one of the following
conditions is satisfied (the third possibility is only for even
$N$):
\begin{equation}
\label{singular}
 P(1)=0,~~~~2s_1+N=0,~~~~2s_{\frac N2-1}+Ns_{\frac N2}=0.
\end{equation}
Singular curves have singularities of the generalized conifold
type, $\xi\eta - \hat y^2+\hat x^n=0$ with $n\ge2$ at one of {\bf
A},{\bf B},{\bf C}. An exception is the case with $N$ even and
$P(x)=(1-x)^{N/2}$ for which the target space becomes an $A_1$
singularity fibered over the $x$-plane, $\xi\eta - \hat y^2=0$.

\subsection{Relation to M-theory curve}

Curves of genus zero describe the moduli space of vacua with
trivial Wilson lines and no low energy $U(1)$ gauge symmetry. The
D5-brane {\bf D} can move on this curve freely and so its position
on the curve serves as a good coordinate on this branch of moduli
space. In other words, the $g=0$ branch of moduli space is
identified with the $g=0$ curve. Here we would like to relate
these $g=0$ curves with the moduli spaces obtained earlier from
M-theory.

We first relate the functions $\eta_{1,2,3}$ on the $g=0$ branch
found in the M-theory analysis to various disc instanton factors
of the type IIA GLSM picture, and then to coordinates $x,y$ of the
IIB picture. The three classical $g=0$ phases were described in
IIA by a D6-brane ending on one of the three legs of the toric
skeleton.

Phase {\sf 1} corresponds to the D6-brane ending at $|z_1|^2=c,
z_2=z_3=0$, and similarly for the other phases. In these three
phases the disc instantons take the form
\begin{equation}
({\sf 1})~:~\{|z_1|^2\le c\}/\mathZ_2,~~~
({\sf 2})~:~\{|z_2|^2\le c\}/\mathZ_N,~~~
({\sf 3})~:~
\left\{\begin{array}{ll}
       \{|z_3|^2\le c\}/\mathZ_{ N}&(N~{\rm even}),\\
       \{|z_3|^2\le c\}/\mathZ_{2N}&(N~{\rm odd}).
\end{array}\right.
\end{equation}
They are related to the lens spaces $Q'_{1,2,3}$ of (\ref{Q'})
upon lifting back to M-theory.
Using the correspondence between GLSM and LG coordinates
$Y_i \leftrightarrow |z_i|^2$ we find $y\sim V(Q'_1)$ around
the large volume point {\sf 1} on the quantum moduli space,
and $x\sim V(Q'_2)$ around the point {\sf 2}.
By comparing with the table \ref{table:DQ} we find
\begin{equation}
  \eta_1 = x,~~~
  \eta_2 = y^{-1},~~~
  \eta_3 =
\left\{\begin{array}{ll}
       s_{N/2}\cdot yx^{-N/2} &(N~{\rm even})\\
       y^2x^{-N}        &(N~{\rm odd}).
\end{array}\right.
\end{equation}
The coefficient $s_{N/2}$ in the expression for $\eta_3$ is
necessary so that $\eta_3=1$ at the point {\sf 3}.

One can translate the parametric representation (\ref{eta-z}) of
the M-theory curve for $g=0$ to type IIB variables using the
relations above, and one obtains a curve of the form
(\ref{curve2}) with
\begin{equation}
 P(x) ~=~ \frac12\left\{(1+x^{1/2})^{N}+(1-x^{1/2})^{N}\right\}
      ~=~ \sum_{j=0}^{[N/2]}\frac{N!x^j}{(2j)!(N-2j)!}.
\label{P-g0}
\end{equation}
For even $N$ the fermion anomaly argument requires $s_{N/2}=+1$.

Our analysis of the asymptotics showed that there are two families
of curves labeled by $s_{N/2}=\pm1$ when $N$ is even.
The moduli space of vacua may consist of several $g=1$ branches
some of which are with $s_{N/2}=1$ and the other are with $s_{N/2}=-1$,
but only those with $s_{N/2}=1$ can be connected to the $g=0$ branch.
It would be interesting to find out what the label
$s_{N/2}$ corresponds to in the previous pictures.

\subsection{Branch structure}

The defining equation for curves of reduced genus $g$ should look like
\begin{equation}
\begin{array}{lrcl}
 (N=2n)  & \hat y^2 ~=~ P_n(x)^2 - (1-x)^N
  &=& G_{2g+1}(x)H_{n-g-1}(x)^2, \\
 (N=2n+1)& \hat y^2 ~=~ P_n(x)^2 - (1-x)^N
  &=& G_{2g+1}(x)H_{n-g}(x)^2,
\end{array}
\label{degen}
\end{equation}
where $P(0)=1$. Here and in the following we put degrees of
polynomials as the suffix. The highest coefficient of $P(x)$ has
to be $\pm1$ when $N$ is even, but it is automatic from the above
ansatz. We will refer to the curve with double roots removed,
$\tilde y^2=G_{2g+1}(x)$, as the {\it reduced curve}.

We have seen that the reduced curves of $g=0$ satisfy the F-term
conditions no matter where $\bf D$ is, as its location is just
related to the values of monopole condensates. So the $g=0$ branch
of the moduli space is the $g=0$ curve itself (not the reduced
curve). On the other hand, on $g=1$ branches the complex structure
of $\Sigma$ is determined by the position of $\bf D$ so that there
exists a meromorphic function with poles at $\bf A,B$ and zeroes
at $\bf C,D$. The location of $\bf D$ and the moduli of a genus
one curve are related in the following way. Suppose {\bf C} is at
$(x,\tilde y)=(\infty,\infty)$ on the reduced curve $\tilde y^2 =
G_3(x)$. Then one can define a map from the reduced curve to its
Jacobian variety
\[
 \mu_{\bf C}({\bf X})=\int_{\bf C}^{\bf X}\frac{dx}{\tilde y},
\]
in terms of which the F-term condition can be expressed as
\[
 \mu_{\bf C}({\bf D}) = \mu_{\bf C}({\bf A}) + \mu_{\bf C}({\bf B}).
\]
The condition for this to hold is
\begin{equation}
\left|\begin{array}{rrr}
 1 & x_{\bf A} &  \tilde y_{\bf A} \\
 1 & x_{\bf B} &  \tilde y_{\bf B} \\
 1 & x_{\bf D} & -\tilde y_{\bf D}
\end{array}\right|
~=~ 0.
\end{equation}

On the $g=1$ branches we shall use the notation $(x_{\bf D},y_{\bf
D})$ to label the position of the brane whilst on the $g=0$
branches we shall use the notation $(x,y)$ since the brane can be
anywhere on the curve. When we approach certain boundaries of the
moduli space we will be able to read off corresponding gauge
dynamics.

The moduli space has boundaries when the brane (at {\bf D}) approaches
one of the punctures ${\bf A}$, ${\bf B}$ or ${\bf C}$ and also when
some of the moduli of the curve approach infinity. 

If the brane approaches ${\bf A}$ then this corresponds to the
$\mathZ_2$-orbifold of deformed conifold with $\mathbb{RP}^3$
growing large in the original IIA setup.
When this happens on a $g=0$ branch we expect to find $SU(N)$ gauge
dynamics whilst on a $g=1$ branch we should find $SU(N)$ broken by a
Wilson line. 

If ${\bf D}$ approaches ${\bf B}$ or ${\bf C}$ this corresponds to the
orbifold of resolved conifold with a large $S^2$ in the original
IIA framework with large $\mathbb{P}^1$.
As we have discussed this can lead to $SU(2)$ or trivial gauge dynamics
from the $\mathbb{R}^4/\mathbb{Z}_2$ singularity. We shall see that such
limits of moduli space can also occur on the $g=1$ branches and
correspond to $SU(2)$ broken by Wilson lines. The possibility of
including such Wilson lines is easy to understand from the M-theory
picture where the $SU(2)$ theory is realized on $S^3/\mathbb{Z}_N$.

Finally, if some of the moduli of the curve approach infinity on a $g=1$
branch, we expect this to describe a limit in which $\mathbb{P}^1 \times
\mathbb{P}^1$ grows large in the original IIA setup.

For example on the $g=0$ branch, let the brane approach the point
${\bf A}$. Since the $g=0$ branch has no low-energy $U(1)$ gauge
dynamics, the $SU(N)$ dynamics should be encoded near the puncture
{\bf A}. In this region, the curve becomes approximately
\begin{equation}
  2P(1)y=(1-x)^N.
\end{equation}
$P(1)$ is nonzero, by equation (\ref{singular}) and one can regard
this as expressing the $SU(N)$ dynamics by identifying $y$ with
the SYM scale and $1-x$ as the gaugino condensate.  One can read
off the SYM dynamics from other punctures in a similar way as long
as the curve is not singular there.

Let us now analyze the cases of small $N$ one by one, starting with $N=1$.

\paragraph{N=1}

In this case $P(x)$ is of degree zero, so it is a constant
$P(x)=P(0)=1$. We have only one $g=0$ branch given by
\begin{equation}
 \hat y^2 ~=~ (y-1)^2 ~=~ x.
\end{equation}
This case is better understood by going back to M-theory and
choosing the M-theory circle from the second $SU(2)$ factor $g_2$.
The semiclassical region {\sf 2}, which is centered at the
puncture {\bf B}, then corresponds to two D6-branes wrapped on a
large $S^3$ of the deformed conifold. The phases ${\sf 1,3}$ are
centered at {\bf A},{\bf C} and correspond to the resolved
conifold with a large $S^2$ and flux.

\paragraph{N=2}

In this case there are two sign choices for $P_1(x)$, $P_1(x)=1\mp x$.
Correspondingly there are two curves of genus zero,
\begin{equation}
\begin{array}{cl}
 (-) & \hat y^2 ~=~ (y-1+x)^2~=~0, \\
 (+) & y^2 - 2y(1+x) + (1-x)^2 ~=~ 0.
\end{array}
\end{equation}
For the $(-)$ choice the IIB target space becomes just the line of
$A_1$ singularity. The other choice agrees with (\ref{P-g0}), and
we claim that it describes the quantum moduli space of vacua for
$N=2$.

\paragraph{N=3}
In this case $P_1(x)=1+sx$ has one modulus, and we have
a one-parameter family of curves of $g=1$,
\begin{equation}
 \hat y^2
 ~=~ (1+sx)^2 - (1-x)^3
 ~=~ x^3+(s^2-3)x^2+(2s+3)x.
\end{equation}
The three punctures {\bf A},{\bf B},{\bf C} are at
\[
 (x_{\bf A},\hat y_{\bf A})=(1,-1-s),~~~
 (x_{\bf B},\hat y_{\bf B})=(0,0  ),~~~
 (x_{\bf C},\hat y_{\bf C})=(\infty,\infty),
\]
The position of the free D5-brane {\bf D} is easily obtained
\begin{equation}
 x_{\bf D}      =2s+3,~~~
 \hat y_{\bf D} =(s+1)(2s+3),~~~
 y_{\bf D}      = 4(s+1)^2.
\end{equation}
This genus one curve should be identified with the branch
$(N_+,N_-)=(1,2)$. From the relation between the moduli of the
curve and the K\"ahler parameters of the GLSM, one finds that the
local $\mathW\mathP^2_{3,1,2}$ becomes large in the IIA side as
$s\to\infty$ and {\bf D} approaches {\bf C}. The curve degenerates
to $g=0$ for the following values of $s$ and $x_{\bf D}$;
\begin{equation}
\begin{array}{llrcl}
 s=  3      , & (x_{\bf D},y_{\bf D})=(9,64),~ & \hat y^2 &=& x(x+3)^2, \\
 s= -1      , & (x_{\bf D},y_{\bf D})=(1,0) ,~ & \hat y^2 &=& x(x-1)^2, \\
 s= -\frac32, & (x_{\bf D},y_{\bf D})=(0,1) ,~ & \hat y^2 &=&
x^2(x-\frac34).
\end{array}
\end{equation}
The first choice of $s$ agrees with (\ref{P-g0}), so it should
correspond to the branch with trivial Wilson line. The
semiclassical points {\sf 1,2,3} of figure \ref{fig:gzero} are
identified with $x_{\bf D}=1,0,\infty$ or {\bf D} approaching one
of the punctures. The other two curves of genus zero are singular,
one at $(x,y)=(1,0)$ and the other at $(0,1)$, as explained in the
previous subsection. They both correspond to {\bf D} approaching
the punctures {\bf A} and {\bf B}, and are at the boundary of the
$g=1$ branch.

The geometric transition involving a non-trivial Wilson line on
the D6-branes wrapped on $\mathR\mathP^3$ should be described by
the family of $g=1$ curves. In particular, the coordinates
$(x_{\bf D}, y_{\bf D})$ are related to the disc instanton factors
in the GLSM picture in the $g=1$ branch as well. They obey the
equation
\begin{equation}
  y_{\bf D} ~=~ (1-x_{\bf D})^2.
\end{equation}
Near the boundary $(x_{\bf D},y_{\bf D})=(1,0)$ one can read off the
$SU(2)\times U(1)$ SYM dynamics corresponding to $SU(3)$ gauge
symmetry with a nontrivial $\mathZ_2$ Wilson line.
The other boundary $(x_{\bf D},y_{\bf D})=(0,1)$ corresponds to
the classical $SU(2)$ gauge symmetry broken by a $\mathZ_3$
Wilson line.

\paragraph{N=4}

In this case there are two one-parameter families for $P_2(x)$,
$P_2(x)= 1+sx\pm x^2$.
We will study both choices in detail, and try to read off
the correct number of vacua for each theory.

For $(+)$ choice we have a family of $g=1$ curves
\begin{equation}
  \hat y^2 ~=~ (1+sx+x^2)^2-(1-x)^4
           ~=~ (s+2)\{2x^3 + (s-2)x^2 + 2x\},
\label{N4curve}
\end{equation}
with {\bf A},{\bf B},{\bf C},{\bf D} at
\[
 (x_{\bf A},\hat y_{\bf A})=(1,-s-2),~
 (x_{\bf B},\hat y_{\bf B})=(0,0   ),~
 (x_{\bf C},\hat y_{\bf C})=(\infty,\infty),~~~
 (x_{\bf D},\hat y_{\bf D})=(1, s+2).
\]
This family of $g=1$ curves should describe one of the
$(N_+,N_-)=(2,2)$ branches. Note that $x_{\bf D}=1$ everywhere on
this branch. If we are in the vacua where the values of the
gaugino condensate for the two $SU(2)$
factors are of opposite sign, then the sum of the two  $SU(2)$ gaugino
condensates vanishes irrespective of the value of the gauge coupling. 
This suggests that we interpret $1 - x_{\bf D}$
as the sum of the gaugino condensates. We will find a similar result
when we study $N=5$ at an $SU(2) \times SU(3)$ point.

The $g=1$ curve degenerates to $g=0$ at two points,
\begin{equation}
\begin{array}{llrcl}
 s=  6, & (x_{\bf D},y_{\bf D})=(1,16),~ & \hat y^2 &=& 16x(x+1)^2,\\
 s= -2, & (x_{\bf D},y_{\bf D})=(1,0 ),~ & \hat y^2 &=& 0.
\end{array}
\end{equation}
The first one agrees with the M-theory curve of $g=0$, and it
should be identified with the branch with trivial Wilson line. The
latter is the case where the target space becomes a line of $A_1$
singularities.

For $(-)$ choice we have a family of $g=1$ curves
\begin{equation}
  \hat y^2 ~=~ (1+sx-x^2)^2-(1-x)^4
           ~=~ (-2s+4)x^3 + (s^2-8)x^2 + (2s+4)x,
\end{equation}
with {\bf A},{\bf B},{\bf C},{\bf D} at
\[
 (x_{\bf A},\hat y_{\bf A})=(1,-s),~
 (x_{\bf B},\hat y_{\bf B})=(0,0  ),~
 (x_{\bf C},\hat y_{\bf C})=(\infty,\infty),~~~
 (x_{\bf D},\hat y_{\bf D})=(\frac{2+s}{2-s},\frac{s(2+s)}{2-s}).
\]
One can also find a strange relation between $x_{\bf D}$ and $y_{\bf D}$:
\[
 y_{\bf D} ~=~ -\frac{2s^3}{(2-s)^2}
           ~=~ \frac{(1-x_{\bf D})^3}{1+x_{\bf D}}.
\]
This family degenerates to $g=0$ at three points,
\begin{equation}
\begin{array}{llrcl}
 s=  0, & (x_{\bf D},y_{\bf D})=(1,0)     ,~     & \hat y^2 &=& 4x(x-1)^2,
\\
 s= -2, & (x_{\bf D},y_{\bf D})=(0,1)     ,~     & \hat y^2 &=& 4x^2(2x-1),
\\
 s=  2, & (x_{\bf D},y_{\bf D})=(\infty,\infty),~& \hat y^2 &=& 4x(2-x).
\end{array}
\label{N=4-}
\end{equation}
These degenerations all correspond to the boundary of the moduli
space where ${\bf D}$ approaches one of the three punctures, and
they are all singular. There is therefore no $g=0$ branch
connected to this $g=1$ branch.

Let us count the number of vacua for some semi-classical regions of
the $g=1$ branch where the microscopic theory is a ${\cal N}=1$
SYM theory.
First, near $(x_{\bf D},y_{\bf D})=(1,0)$ the UV theory is
$SU(2)\times SU(2)\times U(1)$ SYM theory which has four vacua.
For each small value of $y_{\bf D}$ there is a single value of
$x_{\bf D}$ on the $(+)$ branch and three values of $x_{\bf D}$
on the $(-)$ branch.
Secondly, near $(x_{\bf D},y_{\bf D})=(0,1)$ the UV theory
should be a seven-dimensional $SU(2)$ gauge theory compactified on
$S^3/\mathZ_4$ with a non-trivial $\mathZ_4$ Wilson line.
Note that such a Wilson line is unique up to conjugations by $SU(2)$.
The corresponding vacuum can be found on the $(-)$ branch and is
the second line of (\ref{N=4-}).
The argument proceeds in the same way for the third semi-classical
region $(x_{\bf D},y_{\bf D})=(\infty,\infty)$, where the UV theory
is again seven-dimensional $SU(2)$ gauge theory on $S^3/\mathZ_4$
with a $\mathZ_4$ Wilson line and the corresponding vacuum is the
third line of (\ref{N=4-}).

\paragraph{N=5}

From here on we have to tune the moduli of the polynomial
$P_2(x)=1+s_1x+s_2x^2$ so that the curve has reduced genus $g\le1$.
There are three one-parameter families of $g=1$ curves,
two of which are singular at the punctures {\bf A} and {\bf B}
respectively.
The remaining one is given by
\begin{eqnarray}
 \hat y^2 &=&
 \left(1+\frac{-s^4-3s^3+s^2+3s+4}{2s}x
        +\frac{-3s^3-3s^2-s+1}{2s^2}x^2\right)^2-(1-x)^5 \nn\\
     &=& (x+s^2+2s)^2G_3(x), \nn\\
G_3(x) &=&
   x^3+\frac{s^6+2s^5-5s^4-5s^2-2s+1}{4s^4}x^2
      +\frac{-s^2+s+1}{s^3}x.
\end{eqnarray}
The semi-classical regions of $g=1$ branches labeled by
$(N_+,N_-)=(1,4)$ and $(3,2)$ are identified with
$s\sim 0$ and $s\sim\infty$, respectively.
This shows that the two classical $g=1$ branches are
on the same branch of quantum moduli space.
On the reduced curve $\tilde y^2=G_3(x)$ the three punctures
are located at
\begin{equation}
 (x_{\bf A},\tilde y_{\bf A})=(1,\frac{(s-1)(s+1)^2}{2s^2}),~
 (x_{\bf B},\tilde y_{\bf B})=(0,0),~
 (x_{\bf C},\tilde y_{\bf C})=(\infty,\infty),
\end{equation}
so the free D5-brane sits at
\begin{equation}
  x_{\bf D}=\frac{-s^2+s+1}{s^3},~~~
  \tilde y_{\bf D}=\frac{(s-1)(s+1)^2(s^2-s-1)}{2s^5},~~~
  y_{\bf D}=\frac{(s-1)^4(s+1)^6}{s^8}.
\end{equation}
Generically there are three values of $s$ corresponding to a given
$x_{\bf D}$, so this $g=1$ branch is covering the $x_{\bf D}$-space
three times.
Two of the three sheets meet at
\[
 x_{\bf D}=1~~(s=1,-1,-1),~~~~x_{\bf D}=-\frac5{27}~~(s=-\frac35,3,3).
\]

One finds $(x_{\bf D},y_{\bf D})=(\infty,\infty)$ or
$(0,\infty)$ at the two large-volume points corresponding to
$(N_+,N_-)=(1,4)$ or $(3,2)$.
Aside from them, the reduced curve degenerates to $g=0$ for the
following values of $s$ and $x_{\bf D}$:
\begin{equation}
\begin{array}{ll}
 s= -2\pm\sqrt5, &
 (x_{\bf D},y_{\bf D})=(45\pm20\sqrt5,62976\pm28160\sqrt5),~~~
  P_2(x) = 1+10x+5x^2, \\
 s= -1 &
 (x_{\bf D},y_{\bf D})=(1,0),~~~
  P_2(x) = (1-x)^2, \\
 s=  1 &
 (x_{\bf D},y_{\bf D})=(1,0),~~~
  P_2(x) = (1+3x)(1-x), \\
 s= \frac12(1\pm\sqrt5), &
 (x_{\bf D},y_{\bf D})=(0,1),~~~
  P_2(x) = 1-\frac52x-\frac54(1\pm\sqrt5)x^2.
\end{array}
\end{equation}
The first of these agrees with the M-theory $g=0$ curve and
corresponds to the branch with trivial Wilson line. Interestingly,
the $g=1$ branch meets this $g=0$ branch at two different points.
Near $s=\pm1$ on $g=1$ branch, $(x_{\bf D},y_{\bf D})$ approaches
$(1,0)$, so one should be able to read off the SYM dynamics from
\[
  y_{\bf D}   = \frac{(s+1)^6(s-1)^4}{s^8},~~~~
  1-x_{\bf D} = \frac{(s+1)^2(s-1)}{s^3}.
\]
Near $s=1$ one can read off the $SU(4)$ dynamics from the
approximate relation $4y_{\bf D} \simeq (1-x_{\bf D})^4$. The
situation is more interesting at $s=-1$. In this case $y_{\bf
D}\sim 16(s+1)^6$ and $1-x_{\bf D}\sim 2(s+1)^2$ to leading order,
so one can read off the $SU(3)$ dynamics at this order. Including
the next-to leading order terms one finds
\[
 1-x_{\bf D} ~\simeq~
  2\left(\frac{y_{\bf D}}{16}\right)^{\frac26}
 + \left(\frac{y_{\bf D}}{16}\right)^{\frac36},
\]
so the six vacua of $SU(3)\times SU(2)$ SYM theory can be read
off correctly.
Indeed this agrees with our earlier interpretation of $1-x_{\bf D}$
as a sum of gaugino condensates for the two gauge groups.
Finally, near $s=\frac12(1\pm\sqrt5)$ one obtains approximately
linear relations between $1-y_{\bf D}$ and $x_{\bf D}$, describing
the $SU(2)$ gauge symmetry broken by non-trivial $\mathZ_5$
Wilson lines.
The two values of $s$ will correspond to two physically different
Wilson lines, $W=\exp(\frac{2\pi i\tau_3}{5})$
and  $W=\exp(\frac{4\pi i\tau_3}{5})$.

\paragraph{N=6}

Let us first discuss the case with $s_3=+1$,
$P(x)=1+s_1x+s_2x^2+x^3$:
\begin{eqnarray}
  \hat y^2
  &=& (1+s_1x+s_2x^2+x^3)^2-(1-x)^6 \nn\\
  &=& x\{2+(s_1-3)x+(s_2+3)x^2\}\{s_1+3+(s_2-3)x+2x^2\}.
\end{eqnarray}
Since we do not want the degeneration of roots to occur at
$x=0,1,\infty$, we discard the curves of $g=1$ given by
$s_1=-3$, $s_2=-3$ or $s_1+s_2=-2$.
There are thus two families of genus one curves defined by the
conditions
\begin{equation}
 {\rm (I)}~  s_2=\frac18(s_1-3)^2-3~~~~{\rm or}~~~~
 {\rm (II)}~ s_1=\frac18(s_2-3)^2-3.
\end{equation}
We define the reduced curve for each branch as follows:
\begin{equation}
\begin{array}{crcl}
{\rm (I)}   & \tilde y^2 &=& (s_1+3)x\;+(\frac18(s_1-3)^2-6)x^2+2x^3,\\
{\rm (II)}  & \tilde y^2 &=& (s_2+3)x^3+(\frac18(s_2-3)^2-6)x^2+2x.
\end{array}
\end{equation}
The location of the free D5-brane on the curve is
given for each case by
\begin{equation}
\begin{array}{clll}
{\rm (I)} & x_{\bf D}=\ds\frac{s_1+3}{2},&
     \tilde y_{\bf D}=\ds\frac{(s_1+1)(s_1+3)}{4\sqrt2},&
            y_{\bf D}=\ds\frac{(s_1+1)^4}{16},\\
{\rm (II)}& x_{\bf D}=\ds\frac{2}{s_2+3},&
     \tilde y_{\bf D}=\ds\frac{s_2+1}{\sqrt2(s_2+3)},&
            y_{\bf D}=\ds\frac{(s_2+1)^4}{2(s_2+3)^3}.
\end{array}
\end{equation}
On each of these two branches $(x_{\bf D},y_{\bf D})$ obey
\begin{equation}
 ({\rm I}) ~~y_{\bf D}=(1-x_{\bf D})^4,~~~~
 ({\rm II})~~y_{\bf D}=(1-x_{\bf D})^4/x_{\bf D}.
\end{equation}
So there are eight values of $x_{\bf D}$ for each small value of
$y_{\bf D}$, accounting for half the required vacua for the
$SU(N_+)\times SU(N_-)$ SYM theories with $(N_+,N_-)=(4,2)$ and $(2,4)$.
The limit $s_1\to\infty$ or $(x_{\bf D},y_{\bf
D})=(\infty,\infty)$ on (I) corresponds to a semi-classical limit
with large $\mathW\mathP^2_{3,1,2}/\mathZ_2$, and similarly $s_2\to\infty$
or $(x_{\bf D},y_{\bf D})=(0,\infty)$ on (II) corresponds to
another large $\mathW\mathP^2_{3,2,1}/\mathZ_2$.
There are also points $(x_{\bf D},y_{\bf D})=(0,1)$ on (I) and
$(x_{\bf D},y_{\bf D})=(\infty,\infty)$ on (II) corresponding to
two different semi-classical points with $SU(2)$ broken by
a non-trivial $\mathZ_6$ Wilson lines.
However, there are two inequivalent $\mathZ_6$ Wilson lines
in $SU(2)$ and we have recovered only one of them at
each semiclassical point.

Besides these points, further degeneration to $g=0$ occurs at
\begin{equation}
\begin{array}{lllrcl}
 {\rm (I)} &
 s_1=15, &
 (x_{\bf D},y_{\bf D})=(9,4096),&
  P_3(x) &=& 1+15x+15x^2+x^3,\\
 &
 s_1=-1, &
 (x_{\bf D},y_{\bf D})=(1,0),&
  P_3(x) &=& 1-x-x^2+x^3,\\
 {\rm (II)} &
 s_2=15, &
 (x_{\bf D},y_{\bf D})=(\frac19,\frac{4096}{729}),&
  P_3(x) &=& 1+15x+15x^2+x^3,\\
 &
 s_2=-1, &
 (x_{\bf D},y_{\bf D})=(1,0),&
  P_3(x) &=& 1-x-x^2+x^3.
\end{array}
\end{equation}
The curve of genus zero with $s_1=s_2=15$ agrees with the M-theory
curve, so it describes the $g=0$ branch of quantum moduli space.
The two $g=1$ branches (I) and (II) are attached to two different
points on the $g=0$ branch.

Let us next consider the case $s_3=-1$, $P(x)=1+s_1x+s_2x^2-x^3$;
\begin{equation}
 \hat y^2 ~=~ x\{s_1+3 +(s_2-3)x\}\{2+(s_1-3)x +(s_2+3)x^2-2x^3\}
\end{equation}
We first look for curves with $g=1$, excluding singular ones given
by $s_1=-3,~ s_2=3~{\rm or}~s_1+s_2=0$. The degeneration of
branching points of interest occurs only when the third factor in
the right hand side develops a double root. Denoting the double
root by $s$ we immediately obtain
\begin{equation}
 s_1 ~=~ -2s^2+3-\frac4s,~~~~
 s_2 ~=~ 4s-3+\frac{2}{s^2}.
\end{equation}
The location of the free D5-brane is given by
\begin{equation}
 x_{\bf D} = \frac{s+2}{s(2s+1)},~~~~
 y_{\bf D} = \frac{16(1-s)^3(1+s)^5}{s^4(2s+1)^3},~~~~
\label{N=6-}
\end{equation}
The semi-classical points of the $g=1$ branch are identified with
$s=0,\infty$ and $(x_{\bf D},y_{\bf D})=(\infty,\infty)$.
Degeneration to $g=0$ occurs at
\begin{equation}
\begin{array}{lll}
 s=1, &
 (x_{\bf D},y_{\bf D})=(1,0), &
 P_3(x)=(1-x)^3,\\
 s=-1, &
 (x_{\bf D},y_{\bf D})=(1,0), &
 P_3(x)=(1-x)(1+6x+x^2),\\
 s=-2, &
 (x_{\bf D},y_{\bf D})=(0,1), &
 P_3(x)=1-3x-\frac{21}{2}x^2-x^3,\\
 s=-1/2, &
 (x_{\bf D},y_{\bf D})=(\infty,\infty), &
 P_3(x)=1+\frac{21}{2}x+3x^2-x^3.
\end{array}
\end{equation}
They are all singular at one of the punctures. Again, for the
sign choice of $s_3=(-)$ there is no $g=0$ branch connected to
the $g=1$ branch.

Using (\ref{N=6-}) one can easily check that, near
$(x_{\bf D},y_{\bf D})=(1,0)$, there are eight values of $x_{\bf D}$
for each fixed $y_{\bf D}$ (three are near $s=1$ and five are near $s=-1$).
Also, near each of $(x_{\bf D},y_{\bf D})=(0,1)$ and $(\infty,\infty)$
there is one vacuum corresponding to a non-trivial $\mathZ_6$
Wilson line.
All the semi-classical vacua with a $U(1)$ in the IR were thus
identified with the special points on the $g=1$ branch.

\vskip5mm
\noindent{\bf Acknowledgments}
\vskip2mm

We wish to thank 
R.~Dijkgraaf,
J.~Gomis,
A.~Hanany,
M.~Herbst,
T.~Hirayama,
M.~Jinzenji,
S.~Kadir,
K.~Kennaway,
S.~Minabe,
T.~Okuda,
J.~Park,
A.~Peet,
E.~Poppitz,
S.~Rinke,
A.~Takahashi
and especially K.~Hori for useful discussions and comments.
KH was supported in part by NSERC of Canada, and DCP was
funded by PREA of Ontario.

\newpage

\begin{center}{\large\bf References}\end{center}\par


\begin{thebibliography}{999}

\bibitem{Vafa:2000wi}
C.~Vafa,
J.\ Math.\ Phys.\  {\bf 42}, 2798 (2001).

\bibitem{Gopakumar:1998ki}
R.~Gopakumar and C.~Vafa,
Adv.\ Theor.\ Math.\ Phys.\  {\bf 3}, 1415 (1999).

\bibitem{Acharya:2000gb}
B.~S.~Acharya,
arXiv:hep-th/0011089.

\bibitem{Atiyah:2000zz}
M.~Atiyah, J.~M.~Maldacena and C.~Vafa,
J.\ Math.\ Phys.\  {\bf 42}, 3209 (2001).

\bibitem{Atiyah:2001qf}
M.~Atiyah and E.~Witten,
Adv.\ Theor.\ Math.\ Phys.\  {\bf 6}, 1 (2003).

\bibitem{Friedmann:2002ct}
T.~Friedmann,
Nucl.\ Phys.\ B {\bf 635}, 384 (2002).

\bibitem{Ita:2002ws}
H.~Ita, Y.~Oz and T.~Sakai,
JHEP {\bf 0204}, 001 (2002).

\bibitem{Cachazo:2002zk}
F.~Cachazo, N.~Seiberg and E.~Witten,
JHEP {\bf 0302}, 042 (2003).

\bibitem{Cachazo:2003yc}
F.~Cachazo, N.~Seiberg and E.~Witten,
JHEP {\bf 0304}, 018 (2003).

\bibitem{Aganagic:2002wv}
M.~Aganagic, A.~Klemm, M.~Marino and C.~Vafa,
JHEP {\bf 0402}, 010 (2004).

\bibitem{Okuda:2004rg}
T.~Okuda and H.~Ooguri,
Nucl.\ Phys.\ B {\bf 699}, 135 (2004).

\bibitem{Aganagic:2001jm}
M.~Aganagic and C.~Vafa,
JHEP {\bf 0305}, 061 (2003).

\bibitem{Brandhuber:2001yi}
A.~Brandhuber, J.~Gomis, S.~S.~Gubser and S.~Gukov,
Nucl.\ Phys.\ B {\bf 611}, 179 (2001).

\bibitem{PandoZayas:2001iw}
L.~A.~Pando Zayas and A.~A.~Tseytlin,
Phys.\ Rev.\ {\bf D63}, (2001).

\bibitem{Cvetic:2001kp}
M.~Cvetic, G.~W.~Gibbons, H.~Lu and C.~N.~Pope,
Phys.\ Lett.\ B {\bf 534}, 172 (2002).

\bibitem{Leung:1997tw}
N.~C.~Leung and C.~Vafa,
Adv.\ Theor.\ Math.\ Phys.\  {\bf 2}, 91 (1998).

\bibitem{Aganagic:2000gs}
M.~Aganagic and C.~Vafa,
arXiv:hep-th/0012041.

\bibitem{Aganagic:2003db}
M.~Aganagic, A.~Klemm, M.~Marino and C.~Vafa,
arXiv:hep-th/0305132.

\bibitem{Aganagic:2001nx}
M.~Aganagic, A.~Klemm and C.~Vafa,
Z.\ Naturforsch.\ A {\bf 57}, 1 (2002).

\bibitem{Griffiths-Harris}
P.~Griffiths and J.~Harris,
``Principles of Algebraic Geometry,''
John Wiley \& sons, Inc.

\bibitem{Hori:2000kt}
K.~Hori and C.~Vafa,
arXiv:hep-th/0002222.

\bibitem{Hori:2000ck}
K.~Hori, A.~Iqbal and C.~Vafa,
arXiv:hep-th/0005247.

\bibitem{Strominger:1995cz}
A.~Strominger,
Nucl.\ Phys.\ B {\bf 451}, 96 (1995).

\bibitem{Greene:1995hu}
B.~R.~Greene, D.~R.~Morrison and A.~Strominger,
Nucl.\ Phys.\ B {\bf 451}, 109 (1995).

\end{thebibliography}
\end{document}